%% file: ms.tex
\PassOptionsToPackage{prologue,dvipsnames}{xcolor} 
\PassOptionsToPackage{hyphens}{url} 
\documentclass[acmsmall,screen,noacm]{acmart}
\renewcommand\footnotetextcopyrightpermission[1]{} 
\setcopyright{none}
\input{includes}

\input{macros}

\title{Nonmalleable Progress Leakage}
\author{Ethan Cecchetti}
\orcid{0000-0001-7900-8328}
\affiliation{
  \institution{University of Wisconsin--Madison}
  \city{Madison}
  \state{Wisconsin}
  \country{USA}
}
\authornote{Work done in part while author was at the University of Maryland.}
\email{cecchetti@wisc.edu}
\date{}

\input{rule-defs}

\begin{document}

\input{abstract}

\maketitle
\thispagestyle{empty}
\fancyfoot{}

\input{intro}
\input{label-model}
\input{security-defs}

\input{calculus}

\input{inference}
\input{rocq-overview}
\input{related}
\input{conclusion}
\input{acks}

\bibliography{../../ethan,../ephemeral}

\appendix
\input{full-calculus}

\end{document}

%% file: includes.tex
\usepackage{microtype}
\usepackage{mathpartir}
\usepackage{xspace}
\usepackage{xparse}
\usepackage{suffix}
\usepackage{mathtools}
\usepackage{scalerel}
\usepackage{outlines}
\usepackage{listings}
\usepackage{cprotect}
\usepackage{enumitem}

\usepackage{tikz}
\usetikzlibrary{arrows,arrows.meta,calc,fit,matrix,positioning,shapes.misc,shapes,patterns}

\usepackage{amsmath,amsthm,amssymb}
\usepackage[noline]{algorithm2e}

\usepackage{langrules}
\SetRuleLabelLoc{left}

\theoremstyle{plain}
\newtheorem{theorem}{Theorem}
\newtheorem{proposition}{Proposition}
\newtheorem{lemma}{Lemma}
\newtheorem{corollary}{Corollary}
\theoremstyle{definition}
\newtheorem{definition}{Definition}

%% file: macros.tex
\makeatletter
\renewcommand{\paragraph}[1]{%
  \@startsection{paragraph}{4}{\z@}%
                {1ex plus 0.25ex minus 0.25ex}{-1ex plus -0.25ex minus -0.5ex}%
                {\normalfont\upshape\normalsize\bfseries}*{#1.}%
}
\makeatother

\definecolor{codeBlue}{rgb}{0,0,0.85}
\definecolor{orangeGold}{rgb}{0.72, 0.4, 0}
\definecolor{forestGreen}{rgb}{0,0.5,0}
\definecolor{maroon}{rgb}{0.7,0,0}
\lstdefinelanguage{examplelang}{
  keywords=[1]{untrusted, user, input, server, mapper, appCode},
  keywordstyle=[1]\color{orangeGold},
  keywords=[2]{trusted, system},
  keywordstyle=[2]\color{codeBlue},
  keywords=[3]{output, display, render},
  keywordstyle=[3]\color{Plum},
  sensitive=true,
  escapeinside={(*}{*)},
  morecomment=[l]{//}, 
  commentstyle=\color{ForestGreen},
}
\newlength{\lstmargin}
\setbox1=\hbox{\ttfamily\footnotesize{1}}
\newlength{\linenumgap}
\newcommand{\programfont}[1]{\ensuremath{\mathsf{#1}}\xspace}

\newcommand{\DefineEvalRule}[3]{\DefineRule[E#1Rule]{E-#1}{#2}{#3}}
\newcommand{\DefineTypeRule}[3]{\DefineRule[T#1Rule]{#1}{#2}{#3}}

\newcommand{\Vars}{\ensuremath{\mathcal{V}}}
\newcommand{\pc}{\ensuremath{\mathit{pc}}\xspace}
\newcommand{\nt}{\mathit{nt}}
\newcommand{\st}{\mathit{st}}
\newcommand{\Labs}{\ensuremath{\mathcal{L}}}
\newcommand{\ConfLabs}{\mathcal{C}}
\newcommand{\IntegLabs}{\mathcal{I}}
\newcommand{\NN}{\mathbb{N}}
\newcommand{\adv}{\mathcal{A}}
\newcommand{\Pub}{\mathcal{P}}
\newcommand{\Trust}{\mathcal{T}}
\newcommand{\D}{\mathcal{D}}
\newcommand{\T}{T}
\newcommand{\TT}{\mathbb{T}}
\newcommand{\trc}{t}
\newcommand{\pfx}{p}
\newcommand{\evtseq}{s}
\newcommand{\Input}{\operatorname{in}}
\newcommand{\In}{\sigma}

\makeatletter
\newsavebox{\@hgbox}
\newcommand{\@hourglass}{
  \savebox{\@hgbox}{%
      \setlength{\unitlength}{0.37cm}
      \begin{picture}(1.0,1.2)%
      \thicklines%
      \put(0,1.1){\line(2,-3){0.8}}%
      \put(.05,1.1){\line(2,-3){0.8}}%
      \put(0,-0.1){\line(2,3){0.8}}%
      \put(0.80,1.1){\line(-1,0){0.8}}%
      \put(0.85,-0.1){\line(-1,0){0.85}}%
      \end{picture}%
  }%
  \operatorname{\scalerel*{\usebox{\@hgbox}}{\ensuremath{X}}}%
}
\newcommand{\reflect}{\@hourglass}
\makeatother
\newcommand{\voice}{\nabla}
\newcommand{\view}{\Delta}

\newcommand{\proves}{\vdash}
\newcommand{\tpout}{\diamond}

\newcommand{\flowsto}{\sqsubseteq}
\newcommand{\nflowsto}{\not\sqsubseteq}
\newcommand{\cflowsto}{\flowsto_\ConfLabs}
\newcommand{\iflowsto}{\flowsto_\IntegLabs}

\newcommand{\join}{\sqcup}
\newcommand{\meet}{\sqcap}
\newcommand{\lequiv}[1][\D]{\lequiv*{}{#1}}
\WithSuffix\newcommand\lequiv*[2]{\approx^{#1}_{#2}}

\newcommand{\lequivG}[1][\D]{\lequiv*{\Gamma}{#1}}

\newcommand{\Mequiv}[1][\D]{\cong^\Gamma_{#1}}

\newcommand{\Tequiv}[1][\D]{\approxeq^\Gamma_{#1}}

\newcommand{\nmifEqIn}[1][(\Pub,\Trust)]{\nmifEqIn*{}{#1}}

\WithSuffix\newcommand\nmifEqIn*[2]{%
  \operatorname{\mathit{nmif-eq-in}}%
  \if\relax\detokenize{#1}\relax\else^{#1}\fi%
  \if\relax\detokenize{#2}\relax\else_{#2}\fi%
}

\newcommand{\Silent}[1][\D]{\mathit{Sil}^\Gamma_{#1}}

\newcommand{\from}{\leftarrow}
\newcommand{\alt}{\mkern3mu\mid\mkern3mu}

\newcommand{\dom}{\operatorname{dom}}
\newcommand{\defeq}{%
  \mathchoice{\hspace{0.5ex minus 0.5ex}\overset{\raisebox{0pt}[0.5ex]{\normalfont\tiny def}}{=}\hspace{0.5ex minus 0.5ex}}
             {\overset{\raisebox{0pt}[0.5ex]{\normalfont\tiny def}}{=}}
             {\overset{\raisebox{0pt}[0pt]{\normalfont\tiny def}}{=}}
             {\overset{\raisebox{0pt}[0pt]{\normalfont\tiny def}}{=}}%
}
\newcommand{\defIff}{\defeq}
\newcommand{\e}{\epsilon}

\makeatletter
\newcommand{\raw@subst}[2]{
  \def\subst@inner##1##2##3\_nil{##1 \mapsto ##2\ifx\relax##3\relax\else,\mkern2mu\subst@inner##3\_nil\fi}%
  \mathchoice{#1{\left[\subst@inner#2\_nil\right]}}
             {#1[\subst@inner#2\_nil]}
             {#1[\subst@inner#2\_nil]}
             {#1[\subst@inner#2\_nil]}%
}
\newcommand{\subst}[3]{\raw@subst{#1}{{#2}{#3}}}
\WithSuffix\newcommand\subst*[2]{\raw@subst{#1}{#2}}
\makeatother

\newcommand{\Imp}{\textsc{Imp}\xspace}
\newcommand{\seq}{\mathbin{\mathord{;}}}
\newcommand{\assign}{\coloneqq}
\newcommand{\Skip}{\programfont{skip}}
\newcommand{\Stop}{\programfont{stop}}
\newcommand{\IfN}{\programfont{if}}
\newcommand{\ThenN}{\programfont{then}}
\newcommand{\ElseN}{\programfont{else}}

\newcommand{\ITE}[3]{\IfN~{#1}~\ThenN~{#2}~\ElseN~{#3}}
\WithSuffix\newcommand\ITE*[3]{%
  \begin{array}[t]{@{}l@{~}l@{}}
    \IfN~{#1} & \ThenN~{#2} \\ & \ElseN~{#3}
  \end{array}%
}
\newcommand{\WhileN}{\programfont{while}}
\renewcommand{\While}[2]{\WhileN~{#1}~\programfont{do}~{#2}}
\newcommand{\PDown}{\ensuremath{\operatorname{\programfont{pdown}}}\xspace}

\newcommand{\ctx}[2]{\left\langle{#1}, {#2}\right\rangle}
\newcommand{\minwidthbox}[2]{\makebox[{\ifdim#1<\width\width\else#1\fi}]{#2}}
\newcommand{\stepsone}[1][]{
  \if\relax\detokenize{#1}\relax
    \longrightarrow
  \else
    \xrightarrow{\minwidthbox{1em}{\ensuremath{\scriptstyle#1}}}
  \fi}
\newcommand{\stepsmany}[2][*]{\stepsone[#2]^{#1}}
\newcommand{\bstep}[2][\D]{\curvearrowright^{#1}_{#2}}
\newcommand{\stopEff}{\operatorname{stp}}
\newcommand{\assignEff}{\operatorname{a}}
\newcommand{\termEff}{\operatorname{pd}}
\newcommand{\trace}[2]{(#1, #2)}

\newcommand{\hp}[3]{\ensuremath{\mathbf{#1}^{#2}_{#3}}\xspace}
\newcommand{\hpG}[2]{\hp{#1}{\Gamma}{#2}}
\newcommand{\newHp}[2][\adv]{%
  \expandafter\newcommand\csname #2\endcsname[1][#1]{\hp{#2}{}{##1}}%
  \expandafter\newcommand\csname #2G\endcsname[1][#1]{\hpG{#2}{##1}}%
}
\newHp[\D]{PsNi}
\newHp[\D]{PiNi}
\newHp[\D]{Lfp}
\newcommand{\Prog}[1][\D]{\operatorname{Prog}_{#1}}
\newHp{PiRd}
\newHp{PiTe}
\newHp{PiNmif}
\newHp{Rpl}
\newHp{Tpc}
\newHp{NmPl}
\newHp{PsRd}
\newHp{PsTe}
\newHp{PsNmif}

\newcommand{\Behav}{\operatorname{Behav}}

\newcommand{\inffont}[1]{\mathsf{\color{orangeGold}#1}}
\newcommand{\GetLabelN}{\operatorname{\inffont{elab}}}
\newcommand{\InfRelN}{\operatorname{\inffont{pd-place}}}
\newcommand{\RelSetN}{\operatorname{\inffont{pd-lab-set}}}
\newcommand{\InferN}{\operatorname{\inffont{pd-inf}}}
\SetKw{Assert}{assert}
\usepackage[outline]{contour}
\contourlength{0.2pt}
\SetKwSwitch{Match}{Case}{Default}{match}{with}{case}{otherwise}{endcase}{endmatch}
\SetKwSwitch{Switch}{Pattern}{Other}{switch}{do}{case}{default}{endpattern}{endsw}
\newcommand{\GetLabel}[1][\Gamma]{\GetLabelN_{#1}}
\WithSuffix\newcommand\InfRel*[2][\pc]{\InfRelN_\Gamma({#1}, {#2})}
\WithSuffix\newcommand\RelSet*[2][\pc]{\RelSetN({#1}, {#2})}
\newcommand{\Infer}[2][\pc]{\InferN_\Gamma({#1}, {#2})}

\colorlet{partCol}{codeBlue}
\newcommand{\partColorText}{{\color{partCol}blue}\xspace}
\newcommand{\partSkip}{{\color{partCol}\Skip}}
\newcommand{\partAssign}[2]{{\color{partCol} #1 \mathrel{\widetilde{\assign}} #2}}
\newcommand{\partc}[1][]{{\color{partCol}\tilde{c}_{#1}}}

\newcommand{\seqlab}[1]{\mathrel{\raisebox{0.15ex}{\color{partCol}${;}\scriptstyle\{#1\}$}}}
\newcommand{\WhileLab}[3]{{\color{partCol}\WhileN~{#1}\mathrel{\programfont{do}\raisebox{0.15ex}{$\scriptstyle\{#2\}$}}{#3}}}
\newcommand{\ITELab}[4]{{\color{partCol}\IfN\raisebox{0.15ex}{$\scriptstyle\{#2\}$}~{#1}~\ThenN~{#3}~\ElseN~{#4}}}
\newcommand{\PartPDown}{\programfont{\color{partCol}pdown}}

\newcommand{\erase}[1]{\mathchoice%
  {\left\lfloor #1 \right\rfloor}
  {\lfloor #1 \rfloor}
  {\lfloor #1 \rfloor}
  {\lfloor #1 \rfloor}%
}
\newcommand{\pdle}{\preccurlyeq_\textnormal{\scshape pd}}
\newcommand{\pdeq}{\equiv_\textnormal{\scshape pd}}

\newcommand{\rocqfont}[1]{\ensuremath{\texttt{#1}}\xspace}
\newcommand{\option}{\rocqfont{option}}

%% file: rule-defs.tex
\DefineEvalRule{Stop}{ }{\ctx{\Skip}{\sigma} \stepsone[\stopEff] \ctx{\Stop}{\sigma}}

\DefineEvalRule{Assign}{
  \ctx{e}{\sigma} \Downarrow n \\
  x \in \dom(\sigma)
}{\ctx{x \assign e}{\sigma} \stepsone[\assignEff(x, n)] \ctx{\Skip}{\subst{\sigma}{x}{n}}}

\DefineEvalRule{SeqStep}{
  \ctx{c_1}{\sigma} \stepsone[\alpha] \ctx{c_1'}{\sigma'} \\
  c_1' \neq \Stop
}{\ctx{c_1 \seq c_2}{\sigma} \stepsone[\alpha] \ctx{c_1' \seq c_2}{\sigma'}}

\DefineEvalRule{SeqSkip}{ }{\ctx{\Skip \seq c}{\sigma} \stepsone[\bullet] \ctx{c}{\sigma}}

\DefineEvalRule{IfN}{
  \ctx{e}{\sigma} \Downarrow n \\
  n \neq 0
}{\ctx{\ITE{e}{c_1}{c_2}}{\sigma} \stepsone[\bullet] \ctx{c_1}{\sigma}}

\DefineRule[EIfZRule]{E-If0}{
  \ctx{e}{\sigma} \Downarrow 0
}{\ctx{\ITE{e}{c_1}{c_2}}{\sigma} \stepsone[\bullet] \ctx{c_2}{\sigma}}

\DefineEvalRule{While}{ }{\ctx{\While{e}{c}}{\sigma} \stepsone[\bullet] \ctx{\ITE{e}{(c \seq \While{e}{c})}{\Skip}}{\sigma}}

\DefineEvalRule{PDownStep}{
  \ctx{c}{\sigma} \stepsone[\alpha] \ctx{c'}{\sigma'} \\
  c' \neq \Stop
}{\ctx{\PDown_\ell c}{\sigma} \stepsone[\alpha] \ctx{\PDown_\ell c'}{\sigma'}}

\DefineEvalRule{PDownSkip}{ }{\ctx{\PDown_\ell \Skip}{\sigma} \stepsone[\termEff(\ell)] \ctx{\Skip}{\sigma}}

\DefineTypeRule{Skip}{ }{\Gamma; \pc \proves \Skip \tpout \nt}

\DefineTypeRule{Assign}{
  \Gamma(x) = \ell \\
  \Gamma \proves e : \ell \\
}{\Gamma; \ell \proves x \assign e \tpout \nt}

\DefineTypeRule{If}{
  \Gamma \proves e : \pc \\\\
  \Gamma; \pc \proves c_1 \tpout \nt \\
  \Gamma; \pc \proves c_2 \tpout \nt
}{\Gamma; \pc \proves \ITE{e}{c_1}{c_2} \tpout \nt}

\DefineTypeRule{Seq}{
  \Gamma; \pc_1 \proves c_1 \tpout \nt_1 \\
  \Gamma; \pc_2 \proves c_2 \tpout \nt_2 \\\\
  \pc_1 \flowsto \pc_2 \\
  \nt_1 \flowsto \pc_2 \\
  \nt_1 \flowsto \nt_2
}{\Gamma; \pc_1 \proves c_1 \seq c_2 \tpout \nt_2}

\DefineTypeRule{While}{
  \Gamma \proves e : \pc \\
  \Gamma; \pc \proves c \tpout \pc
}{\Gamma; \pc \proves \While{e}{c} \tpout \pc}

\DefineTypeRule{PDown}{
  \Gamma; \pc \proves c \tpout \nt \\\\
  \nt \flowsto \reflect(\nt) \\
  \pc \flowsto \ell
}{\Gamma; \pc \proves \PDown_\ell c \tpout \ell}

\DefineTypeRule{Variance}{
  \Gamma; \pc' \proves c \tpout \nt' \\\\
  \pc \flowsto \pc' \\
  \nt' \flowsto \nt
}{\Gamma; \pc \proves c \tpout \nt}

%% file: abstract.tex
\begin{abstract}
  Information-flow control systems often enforce progress-insensitive noninterference, as it is simple to understand and enforce.
  Unfortunately, real programs need to declassify results and endorse inputs, which noninterference disallows,
  while preventing attackers from controlling leakage, including through progress channels, which progress-insensitivity ignores.

  This work combines ideas for progress-\emph{sensitive} security with secure downgrading (declassification and endorsement)
  to identify a notion of securely downgrading progress information.
  We use hyperproperties to distill the separation between progress-sensitive and progress-insensitive noninterference
  and combine it with nonmalleable information flow, an existing (progress-insensitive) definition of secure downgrading,
  to define \emph{nonmalleable progress leakage}~(NMPL).
  We present the first information-flow type system to allow some progress leakage while enforcing NMPL,
  and we show how to infer the location of secure progress downgrades.
  All theorems are verified in Rocq.
\end{abstract}

%% file: intro.tex
\section{Introduction}
\label{sec:intro}

Information flow control~(IFC) is a powerful tool for enforcing information security.
The most common guarantee is \emph{noninterference}, which prohibits a program's more-sensitive inputs
from influencing---interfering with---its less-sensitive outputs~\citep{GoguenM82}.
Different requirements give rise to different formulations,
the most popular being \emph{progress-insensitive noninterference}~(PINI)---which allows the program's termination behavior to leak information---
and \emph{progress-sensitive noninterference}~(PSNI)---which does not.
While PSNI provides stronger security, enforcing it requires extreme limits on
any program construct that may not terminate, including simple while loops, leading many tools to enforce PINI instead.

Unfortunately, PINI is unsuitable in many real systems for two opposite reasons.
First, noninterference is too restrictive.
Many applications need untrusted inputs to sometimes influence decisions and allow controlled release of outputs derived using secrets.
Second, the termination channel left open by progress-insensitivity allows arbitrary leakage,
in theory in a single run~\citep{AskarovHSS08}, or easily across multiple runs if the system automatically restarts when hanging.
This leakage makes PINI too permissive in the presence of active attackers.

Applications mixing secrets with code from untrusted sources provide an illustrative example.
\citet{mobileCode12} suggest IFC types to enforce security in such a setting, but nontermination poses a concern.
Consider the pseudocode in Figure~\ref{fig:ex-code} for a mobile app that displays nearby attractions on a map,
revealing only the user's region to the server.
IFC types can ensure that the application-supplied \lstinline{buildMap}
does not send \lstinline{loc} to the server, only \lstinline{region}, but there remain termination channels.
Lines~\ref{ex:lst:loc} or~\ref{ex:lst:region} might hang due to poor signal, revealing precise location details.
Line~\ref{ex:lst:map} could hang benignly, due to a code bug or a non-responsive server,
or maliciously, in an attempt to reveal \lstinline{loc} through a termination channel.

\begin{figure}
  \setbox0=\hbox{\lstinline{   // over them, placing nearby ones on map}\hspace*{\lstmargin}}%
  \centering%
  \cprotect\mbox{\begin{lstlisting}[frame=lines,linewidth=\wd0,backgroundcolor=\color{black!4}]
   loc = system.getLocation()(*\label{ex:lst:loc}*)
   region = system.getRegion()(*\label{ex:lst:region}*)
   // Fetch attractions in region and loop
   // over them, placing nearby ones on map
   render(appCode.buildMap(loc, region))(*\label{ex:lst:map}*)
  \end{lstlisting}\hspace*{\lstmargin}}%
  \caption{Pseudocode for app mapping nearby attractions.}
  \label{fig:ex-code}
\end{figure}

An application might reasonably choose to accept unlikely and minimal leakage due to poor signal
and ignore the app server hanging, which leaks nothing, but disallow the last, malicious, option.
Unfortunately, enforcing PSNI would break the application, disallowing the minimal poor-signal leakage,
while enforcing only PINI would allow the malicious progress leakage.
This work shows how to differentiate these termination channels,
formally define the desired restriction, and enforce it with a type system.

To accomplish this goal, we turn to a long line of work generalizing noninterference
to support different notions of secure declassification and endorsement, known collectively as \emph{downgrading}~\citep{ZdancewicM01,delimitedRelease03,ChongM06,MantelS04,LiZ05,SabelfeldS05,ChongM08,WayeBKCR15,nmifc17,CecchettiYNM21,SolovievBG24}.
Unfortunately, these conditions are generally progress-insensitive,
at best suggesting enforcing a progress-sensitive variant by preventing any progress leakage~\citep{AskarovM11}.
This solution, while effective, imposes the same constraints as PSNI,
either requiring all loop conditions to be fully public values~\citep{VolpanoS97,ONeillCC06},
or prohibiting any publicly visible operation after a loop with a non-public condition~\citep{MooreAC12,BayA20},
making programming nearly untenable and failing to address the above example.

To circumvent this crushing limitation, we allow programs to explicitly downgrade progress information.
Some prior work supports progress downgrades,
but they either appeal to halting oracles~\citep{MooreAC12}, making them unrealistic,
quantify how much information leaks~\citep{MooreAC12}, making them non-compositional,
or provide intensional definitions of security based on the authority of the declassifier~\citep{BayA20}.
Perhaps more importantly, they all consider only confidentiality and are thus unable to create, or even express,
restrictions on who can influence the timing or content of declassifications,
precisely what is needed to safely mix secrets with code and inputs from untrusted sources.

This work addresses this shortcoming by defining progress-based variants of robust declassification~(RD)~\citep{ZdancewicM01},
and its extensions transparent endorsement and nonmalleable information flow~(NMIF)~\citep{nmifc17},
and showing how to restrict progress downgrades to enforce them.
These conditions constrain the impact of attackers on declassifications and secrets on endorsements,
exactly what is needed when combining secrets and untrusted code.
However, all existing enforcement mechanisms are progress-insensitive~\citep{MyersSZ06,ChongM06,AskarovM11,nmifc17,McCallBJ23},
seriously limiting their power to secure complex systems.

To build these conditions, we formalize \emph{leakage-free progress}~(LFP)---the distinction between PSNI and PINI---%
as a hyperproperty~\citep{hyperproperties10}.
Hyperproperties, or sets of sets of traces, provide a framework for relating multiple executions of the same program,
making them ideal for defining complex information security properties.
Our approach draws insights from the definitions of LFP as well as RD and NMIF
to define \emph{nonmalleable progress leakage}~(NMPL),
a progress-based variant of NMIF that separates progress-sensitive and progress-insensitive NMIF.

We also define the first information-flow type system to enforce NMPL without prohibiting all progress leakage.
This result shows that meaningful end-to-end security is achievable in the presence of malicious code and data
without imposing draconian restrictions on basic looping constructs.
For instance, it supports the desired policy in the mapping example above.

Finally, we leverage the static type system to \emph{infer} progress downgrades.
The inference procedure is efficient, adds a minimal number of downgrades,
and is sound and complete with respect to the type system---
any program it produces is well-typed, and it will find a way to add well-typed progress downgrades if one exists.
The existence of such a procedure is a powerful result.
It can automatically verify that all progress leakage satisfies NMPL
while identifying where leakage can occur to support programmer audits.

The main contributions of this work are as follows.
\begin{itemize}[nosep]
  \item Section~\ref{sec:security} defines LFP as a hyperproperty
    and generalizes it to define NMPL and progress-sensitive NMIF.
  \item Section~\ref{sec:calculus} defines a calculus and type system to enforce progress-sensitive NMIF and proves it secure.
  \item Section~\ref{sec:inference} presents a sound, complete, and efficient procedure
    for inferring a syntactically minimal placement of progress downgrade instructions within a program.
  \item All theorems in this paper have been mechanically verified in the Rocq Prover (formerly Coq)~\citep{rocq},
    the first mechanized statements or proofs of RD and NMIF of which we are aware.
    Section~\ref{sec:rocq} gives an overview of the verification effort.
\end{itemize}
Section~\ref{sec:related} presents related work and Section~\ref{sec:conclusion} concludes.

%% file: label-model.tex
\section{Label Model}
\label{sec:label-model}

Before defining new notions of security, we first describe the structure of our security policies.
As is standard, we express information flow policies through a set of labels~$\Labs$ that form a preorder.
That is, there must be a reflexive and transitive ordering relation~$\flowsto$ (pronounced ``flows-to'')
where $\ell_1 \flowsto \ell_2$ means that~$\ell_2$ is at least as restrictive as~$\ell_1$,
so data with label~$\ell_1$ may safely influence data with label~$\ell_2$.

Because the primary security conditions in this work---robust declassification, transparent endorsement, and nonmalleable information flow~(NMIF)---%
concern the interaction of confidentiality and integrity, the labels must express both.
Prior work accomplishes this goal by demanding a distributive lattice over a set of principals representing their authority.
A label is then an ordered pair of the confidentiality authority and the integrity authority~\citep{ChongM06,nmifc17,ZagieboyloSM19,viaduct21}.
This structure is simple and creates an obvious way to convert between confidentiality and integrity,
which is necessary for defining and enforcing security.
However, it also forces the space of confidentiality labels and integrity labels to be the same, limiting its applications.

This work extends NMIF-style definitions and results to a wider range of information-flow policy spaces by
decoupling the confidentiality policies~$\ConfLabs$ from the integrity policies~$\IntegLabs$.
Each must form a preorder (denoted~$\cflowsto$ and~$\iflowsto$, respectively),
but the sets of policies need not be the same, or even contain any overlapping elements.
Intuitively, $c \cflowsto c'$ means the policy~$c'$ demands at least as much secrecy as~$c$,
while $i \iflowsto i'$ means that~$i$ is at least as trusted as~$i'$.
It is therefore safe to use data labeled~$c$ (or~$i$) in a context expecting data labeled~$c'$ (or~$i'$).
Labels are simply ordered pairs $\Labs = \ConfLabs \times \IntegLabs$ with $\flowsto$ defined pointwise.
We also require a least label $\bot \in \Labs$ such that $\bot \flowsto \ell$ for any label~$\ell$.

Though confidentiality policies and integrity policies come from different spaces, we still need a way to relate the two.
We thus require mappings between them in both directions.
Using the language of \citet{nmifc17}, we call these
the \emph{voice}, $\voice : \ConfLabs \to \IntegLabs$, and the \emph{view}, $\view : \IntegLabs \to \ConfLabs$.
The voice of a confidentiality policy~$c$ represents the least trustworthy integrity level such that everyone who can write to~$\voice(c)$ can read~$c$.
Similarly, the view of~$i$ is the most secret confidentiality level where everyone who can write~$i$ can also read~$\view(i)$.

To formalize this intuition, the mappings must satisfy two properties.
First, they must be \emph{anti-monotonic}.
Since $c_1 \cflowsto c_2$ means anyone who can read~$c_2$ can also read~$c_1$,
any integrity level where everyone can read~$c_2$ also guarantees everyone can read~$c_1$.
In other words, $\voice(c_2) \iflowsto \voice(c_1)$.
A dual argument holds for~$\view$.
Second, $i \iflowsto \voice(c)$ means that anyone who can write to data labeled~$i$ must be able to read data labeled~$c$.
But $c \cflowsto \view(i)$ means exactly the same thing!
We therefore require $i \iflowsto \voice(c) \Longleftrightarrow c \cflowsto \view(i)$.
Together, these properties make~$(\voice, \view)$ an \emph{antitone Galois connection}~\citep[see~e.g.,][]{Galatos07}.

As in prior work~\citep{ZagieboyloSM19,viaduct21},
the voice and view combine to form a \emph{reflection} operator~$\reflect : \Labs \to \Labs$
that flips the confidentiality and integrity of a label:
$\reflect(c, i) \defeq (\view(i), \voice(c))$.
This simple construction extends the properties of $\voice$ and~$\view$ to~$\reflect$,
making~$\reflect$ an antitone Galois connection between~$\Labs$ and itself:
$\ell \flowsto \reflect(\ell') \Longleftrightarrow \ell' \flowsto \reflect(\ell)$.

This reflection operator is critical for enforcing downgrade-tolerant security conditions.
Prior work has shown downgrading to be secure only when anyone who can influence the data
can also read it---$i \iflowsto \voice(c)$ for data labeled~$(c, i)$~\citep{nmifc17,ZagieboyloSM19,viaduct21}.
We will see in Section~\ref{sec:calculus} that the same holds for downgrading progress.
Since $(\voice, \view)$ form an antitone Galois connection, this requirement is equivalent to the label flowing to its own reflection.
Using the terminology of \citet{ZagieboyloSM19}, we refer to labels that fail this requirement as \emph{compromised}.
\begin{definition}[Compromised Label~\citep{ZagieboyloSM19}]
  \label{defn:compromised}
  A label~$\ell$ is \emph{compromised} if $\ell \nflowsto \reflect(\ell)$; not everyone who can write to it can read from it.
  A label is \emph{non-compromised} if $\ell \flowsto \reflect(\ell)$.
\end{definition}

\paragraph{Further Generality}
\label{sec:general-labels}
The results of this paper actually hold for an even more general structure than the one described above.
The labels~$\Labs$ need only form an arbitrary preorder, without explicit separation of confidentiality and integrity.
Every pair of labels must have some lower bound, but there need not be a global least element.%
\footnote{If $\Labs$ is finite, pairwise lower bounds are equivalent to a global lower bound, but if $\Labs$ is infinite, pairwise bounds are weaker.}
Finally, the reflection operator~$\reflect$ simply needs to be anti-monotonic, not an antitone Galois connection.

This added generality does not complicate the proofs, but it does complicate intuition.
In fact, I have
yet to find a reasonable intuitive interpretation of the security results with a label model that does not fit the separate confidentiality and integrity structure above.
However, there is no reason to constrain the technical result by the limits of the author's imagination,
so all formal definitions and theorems are stated and proven in the more general model,
except where explicitly noted otherwise.

%% file: security-defs.tex
\section{Progress-Sensitive Hyperproperties}
\label{sec:security}

Progress-(in)sensitive security conditions generalize the classic notion of termination-(in)sensitive security~\citep{VolpanoSI96,VolpanoS97,SabelfeldM03}.
They consider traces of effectful programs and properly account for leakage during execution, even when programs might not terminate.
Existing work on progress-sensitivity has primarily focused on noninterference
and models attacker knowledge by the set of possible initial memories consistent with the attacker's observations~\citep{AskarovS07,AskarovHSS08,AskarovM11,MooreAC12,BayA20}.
Progress-sensitive noninterference~(PSNI) then requires that the set of memories remain constant as the program executes---and the attacker observes more.
Progress-insensitive noninterference~(PINI) is more permissive,
requiring only that an attacker learn no more from observing an event than from knowing the event exists.
This formulation provides a strong intuition for noninterference.
However, its extensions to robust declassification either
treat confidentiality and integrity asymmetrically~\citep{AskarovM11}, making it unclear if it extends to nonmalleable information flow,
or assume traces identify syntactic downgrades~\citep{McCallBJ23},
requiring the language to track leakage for the definition to make sense.%

We instead turn to \emph{hyperproperties}~\citep{hyperproperties10},
another framework commonly used for information security conditions,
including noninterference, robust declassification, and nonmalleable information flow.
A hyperproperty is a set of sets of traces, and various forms of noninterference are classic examples of hyperproperties~\citep[e.g.,][]{hyperproperties10,BeutnerF22,journey-beyond19,SousaD16,UnnoTK21}.
For instance, PSNI---an attacker learning nothing---requires any two traces with public-equivalent inputs to look the same
at every point in execution, lest the adversary distinguish the inputs, thereby learning a secret.
Notably, while many information-security formalizations use hyperproperties, most are not progress sensitive.
They either ignore nonterminating programs entirely---resulting in progress- and termination-insensitive security definitions---
or require termination behaviors to match, but ignore the effects of nonterminating programs,
creating \emph{termination}-sensitive definitions unsuitable for effectful programs.

To formalize progress-sensitive security through hyperproperties, we view program behaviors as traces.
A trace~$\trc$ must include the program's inputs, denoted~$\Input(\trc)$, and any visible effects the program produces during execution.
A trace may be either finite---if the execution terminates---or infinite---if it diverges.
We also require an equivalence relation~$\lequiv$ on both inputs and finite prefixes of traces,
where $\pfx_1 \lequiv \pfx_2$ means~$\pfx_1$ is \emph{indistinguishable} from~$\pfx_2$ to a low observer~$\D$.

Many definitions separate ``low'' from ``high'' by a single label~$L$---%
anything that flows to~$L$ is ``low'' and all other labels are ``high''---%
we follow a more expressive approach and define security relative to an arbitrary downward-closed set of labels~$\D \subseteq \Labs$,
which need not have a maximal element.
Labels in~$\D$ are considered ``low'' while labels not in~$\D$ are ``high.''

These basic building blocks are sufficient to define our security properties.
Section~\ref{sec:enforcing-security} below shows one way to instantiate these primitives and enforce security.

\subsection{Noninterference and Leakage-Free Progress}
\label{sec:ni-lfp}

We begin with noninterference, partially to fill a minor gap in the space of hyperproperty-based noninterference definitions,
but primarily for expository reasons.
The definition of nonmalleable information flow, and thus nonmalleable progress leakage,
builds directly on that of noninterference while adding significant complexity.

The strongest condition, PSNI, is also the simplest to define.
As described above, PSNI requires two traces with indistinguishable inputs to produce indistinguishable executions.
To formalize that as a hyperproperty, recall that a hyperproperty is a set of sets of traces.
$\PsNi$ is then the hyperproperty such that, for any set of traces $T \in \PsNi$ and any pair of traces $\trc_1, \trc_2 \in T$,
if those traces have $\D$-equivalent inputs---$\Input(\trc_1) \lequiv \Input(\trc_2)$---then every observation in one trace
must also appear in the other, up to $\D$-equivalence.
In other words, an adversary who can see only data and events with labels in~$\D$ learns nothing from a full execution trace
beyond what they could learn from that execution's input.

As is standard~\citep{hyperproperties10}, we model observations as finite prefixes of a trace, resulting in the following definition,
where $\pfx \leq \trc$ denotes that~$\pfx$ is a finite prefix of~$\trc$, and~$\TT$ is the set of all possible traces.
\[
  \PsNi \defeq \{ \T \subseteq \TT \mid \forall \trc_1, \trc_2 \in \T\ldotp \Input(\trc_1) \lequiv \Input(\trc_2)
    \implies
  \forall \pfx_1 \leq \trc_1\ldotp \exists \pfx_2 \leq \trc_2\ldotp \pfx_1 \lequiv \pfx_2 \}
\]

Progress-insensitivity, by contrast, allows an attacker to gain information by learning an event exists,
but no additional information from seeing its contents~\citep{AskarovS07}.
The hyperproperty must therefore allow traces to ``stop early'' from $\D$'s perspective due to infinite loops.
When this happens, the trace~$\trc^\uparrow$ with an infinite loop will appear to be a prefix of~$\trc_\downarrow$, the one without.
When considering finite observations, we do not know, a priori, which trace is which.
However, every prefix of~$\trc^\uparrow$ must be indistinguishable from some prefix of~$\trc_\downarrow$,
and any sufficiently short prefix of~$\trc_\downarrow$ must be indistinguishable from a prefix of~$\trc^\uparrow$.%

To make this idea formal,
given prefixes~$\pfx_1$ and~$\pfx_2$ of traces with $\D$-equivalent inputs,
we require one of~$\pfx_1$ and~$\pfx_2$ to be $\D$-equivalent to a prefix of the other.
Letting $\pfx_1 \leq_\D \pfx_2$ denote that~$\pfx_1$ is indistinguishable from a prefix of~$\pfx_2$---%
there is some $\pfx_2' \leq \pfx_2$ such that $\pfx_1 \lequiv \pfx_2'$---%
we define $\PiNi$ as follows.
\[
  \PiNi \defeq \{ \T \subseteq \TT \mid \forall \trc_1, \trc_2 \in \T\ldotp \Input(\trc_1) \lequiv \Input(\trc_2)
    \implies
  \forall \pfx_1 \leq \trc_1, \pfx_2 \leq \trc_2\ldotp (\pfx_1 \leq_\D \pfx_2 \mathrel{\lor} \pfx_1 \geq_\D \pfx_2) \}
\]

\paragraph{Leakage-free progress}
The distinction between PSNI and PINI is whether progress itself can leak information.
A program does not leak information through progress if, after any sequence of events,
the existence of another event reveals no further information about the program's secret inputs.
More generally---including both confidentiality and integrity---%
the sequence of low (public or trusted) events a program produces must entirely determine if another low event occurs.
We call this security property \emph{leakage-free progress}~(LFP) and present the first hyperproperty formalization of it.

To capture the above intuition, begin with the same setup as in PINI:
finite prefixes~$\pfx_1 \leq \trc_1$ and~$\pfx_2 \leq \trc_2$
from traces where $\Input(\trc_1) \lequiv \Input(\trc_2)$.
If~$\pfx_1$ appears to be a \emph{strict} prefix of~$\pfx_2$, denoted $\pfx_1 <_\D \pfx_2$,
that means it is possible for an execution producing the $\D$-visible events in~$\pfx_1$ to visibly progress---produce another $\D$-visible event.
Since LFP demands that the visible events determine the progress behavior,
$\trc_1$~\emph{must} progress, including some $\D$-visible event beyond the end of~$\pfx_1$.
We thus define LFP as follows, letting $\Prog(\pfx, \trc)$ denote that trace~$\trc$ has another $\D$-visible event beyond the end of prefix~$\pfx$.
Formally, $\Prog(\pfx, \trc) \defeq \exists \pfx' \leq \trc\ldotp \pfx <_\D \pfx'$,
\[
  \Lfp \defeq \{ \T \subseteq \TT \mid \forall \trc_1, \trc_2 \in \T, \pfx_1 \leq \trc_1, \pfx_2 \leq \trc_2\ldotp \pfx_1 <_\D \pfx_2 \implies \Prog(\pfx_1, \trc_1) \}
\]
The requirement that $\Input(\trc_1) \lequiv \Input(\trc_2)$ is implicit here;
the premise $\pfx_1 <_\D \pfx_2$ can only hold with indistinguishable inputs.

Notably, the event~$\trc_1$ progresses with need not match that of~$\trc_2$;
LFP allows leakage through the \emph{content} of events, just not their existence.
A program that outputs its secret to a public channel and then terminates satisfies LFP, as progress is not the leakage vector.
On its own, LFP is therefore unlikely to be a useful security condition,
but it precisely defines the distinction between PINI, which allows leakage \emph{only} through progress,
and PSNI, which disallows all leakage.
Indeed, satisfying PSNI is the same as satisfying both PINI and LFP.
\begin{theorem}
  \label{thm:psni-is-pini-lfp}
  For any~$\D \subseteq \Labs$, $\PsNi = \PiNi \cap \Lfp$.
\end{theorem}

These definitions are parameterized for a single attacker,
but each extends simply to \emph{all} attackers by taking the intersection over all downward-closed sets~$\D$:
\begin{mathpar}
  \PsNi[] \defeq \bigcap \PsNi
  \and
  \PiNi[] \defeq \bigcap \PiNi
  \and
  \Lfp[] \defeq \bigcap \Lfp
\end{mathpar}

\paragraph{Hypersafety and Hyperliveness}
\label{sec:ni-safety-and-liveness}

Hyperproperties are often divided into \emph{hypersafety}, requiring that bad things don't happen,
and \emph{hyperliveness}, requiring that good things do happen.
Noninterference is a classic example of not just a hyperproperty, but specifically hypersafety~\citep[e.g.,][]{BeutnerF22,journey-beyond19,SousaD16}.
Indeed, the more common progress-\emph{insensitive} noninterference is hypersafety.

Progress-\emph{sensitive} noninterference, however, is not.
It is subset-closed, but it both prohibits bad things---attackers cannot learn from events they see---
and requires good things---both executions make progress or both silently diverge.
Every hyperproperty is the intersection of a hypersafety property and a hyperlivness property~\citep{hyperproperties10},
and the decomposition of \PsNi[] is precisely progress insensitivity and progress leakage: \PiNi[] is hypersafety, while \Lfp[] is hyperliveness.
This decomposition and a similar note in Section~\ref{sec:nmif-safety-and-liveness}, which are not stated as theorems,
are the only claims in this paper not verified in Rocq.

\subsection{Robust Declassification}
\label{sec:psrd}

Robust declassification~(RD) recognizes that many programs need to declassify information to function, inherently violating noninterference.
Instead, it prohibits an untrusted attacker from influencing the timing or content of declassifications.
Identifying influence requires multiple program executions: two runs with different attacks must produce the same declassifications.
This idea seems to suggest a definition similar to noninterference,
where traces with the same trusted inputs must produce the same declassifications.
Unfortunately, semantically, declassifications are defined as (confidentiality) violations of noninterference,
meaning detecting them requires two executions with different secrets.
Existing formalizations of RD therefore use four runs~\citep{MyersSZ06,ChongM06,nmifc17},
comparing two pairs of inputs where secret inputs differ within a pair
and an attack varies across the pairs.

These definitions model attacks by leaving holes in commands and inserting low-integrity attacker code.
To keep our hyperproperties language-agnostic, we instead vary untrusted values in the initial input.
These formulations produce equivalent guarantees for languages with conditional branches when enforcement theorems quantify over all programs,
as is the case for our example language in Section~\ref{sec:enforcing-security}.
To model a hole with either of two attacks, simply hard code the attacks as branches of an \IfN statement
conditioned on part of the low-integrity input not used elsewhere, and vary only that value.

\paragraph{Defining Attackers}
\label{sec:attackers}
Making these ideas precise requires dividing confidentiality into ``public'' and ``secret'' and integrity into ``trusted'' and ``untrusted.''
Prior approaches that demand confidentiality and integrity be dual policy lattices
require public labels---the attacker's ability to read data---and untrusted labels---the attacker's ability to write data---%
to be the same policy sets~\citep{MyersSZ06,ChongM06,AskarovM11,nmifc17,ZagieboyloSM19,viaduct21,McCallBJ23}.
Since our confidentiality and integrity policies may come from disperate spaces (see Section~\ref{sec:label-model}),
we require a more general definition.
We bound an attacker's power by two downward-closed sets of labels, representing public and trusted policies, respectively,
and require only that the attacker can read any security level they can write;
the view of untrusted integrity must be public.
\begin{definition}[Attacker]
  \label{defn:attacker}
  A pair of label sets $\adv = (\Pub, \Trust)$ is an \emph{attacker} if
  $\Pub = P \times \IntegLabs$ and $\Trust = \ConfLabs \times T$
  for downward-closed sets $P \subseteq \ConfLabs$ and $T \subseteq \IntegLabs$ such that $\view(\overline{T}) \subseteq P$.
\end{definition}
Note that we could have required the voice of secret confidentiality be trusted---$\voice(\overline{P}) \subseteq T$.
The definitions are equivalent since $(\voice, \view)$ form an antitone Galois connection.

Recall from Definition~\ref{defn:compromised} that non-compromised labels---those where $\ell \flowsto \reflect(\ell)$---%
aim to capture labels where anyone who can influence the data can also read it.
That means, for every attacker, a non-compromised label should be either
public---the attacker \emph{can} read it---or trusted---the attacker \emph{cannot} write it---or both.
Indeed, Definition~\ref{defn:attacker} enforces this property.
\begin{proposition}
  \label{prop:non-comp-low}
  For any attacker $(\Pub, \Trust)$ and any label $\ell \in \Labs$, if $\ell \flowsto \reflect(\ell)$,
  then $\ell \in \Pub \cup \Trust$.
\end{proposition}

As with the label model itself (Section~\ref{sec:label-model}), our theorems support a more general definition,
requiring only that~$\Pub$ and~$\Trust$ be downward-closed and satisfy Proposition~\ref{prop:non-comp-low}.
It is again unclear, intuitively, what this more general structure represents.
The proofs, however, rely only on Proposition~\ref{prop:non-comp-low}, so there is no reason to restrict the formalism.

\paragraph{Defining Robust Declassification}
This notion of an attacker allows us to formalize the above intuition for RD
that an attacker should have no influence over the timing or content of declassifications.
Prior definitions only reason about the behavior of terminating traces~\citep{MyersSZ06,ChongM06,nmifc17},
making them not only progress-insensitive, but entirely unable to reason about progress leaks.
Their structure, however, provides useful intuition.
They require that, given any set of four traces $\trc_{11}, \trc_{12}, \trc_{21}, \trc_{22}$ with inputs $\In_{ij} = \Input(\trc_{ij})$,
if
\begin{enumerate}[nosep]
  \item $\In_{11} \lequiv[\Pub] \In_{21}$ and $\In_{12} \lequiv[\Pub] \In_{22}$---$(\In_{11}, \In_{21})$ and $(\In_{12}, \In_{22})$ are the pairs of inputs varying secrets,
  \item $\In_{11} \lequiv[\Trust] \In_{12}$ and $\In_{21} \lequiv[\Trust] \In_{22}$---only the attack varies across the pairs, and
  \item all traces terminate,
\end{enumerate}
then the second attack cannot leak secrets unless the first does as well.
That is, $\trc_{11} \lequiv[\Pub] \trc_{21}$ implies $\trc_{12} \lequiv[\Pub] \trc_{22}$.

This formulation has two major shortcomings.
First, it cannot reason about any divergent programs or constrain progress leakage.
Second, enforcing it requires prohibiting endorsement, as endorsed data may safely influence future declassifications.

The key to solving both problems lies in explicitly considering partial executions
and only restricting declassification prior to any (semantic) endorsements.
More formally, given any program point in the first trace $\pfx \leq \trc_{11}$,
if no (semantic) endorsements have occurred by~$\pfx$ and the first attack cannot yet differentiate the secrets
then the second attack must not leak information to that point.

Checking for semantic endorsements is simple:
only consider prefixes in the second attack $\pfx_{12} \leq \trc_{12}$ when $\pfx_{12} \lequiv[\Trust] \pfx$.
If an endorsement has already occurred, there will be no such prefixes and the condition will hold vacuously.

Checking if the first attack leaks nothing up to~$\pfx$ is more complicated for two reasons.
First, we must pick an appropriate definition for ``leaks nothing.''
One might assume that we should use whatever notion of leakage we aim to constrain:
a PSNI-like structure for constraining all leakage,
a PINI-like structure to allow progress leakage but nothing else,
or an LFP-like structure to prohibit only progress leakage.
Unfortunately, using this approach for both attacks does not give the desired~result.

Consider the following program with a public--untrusted label for~$a$ and secret labels for~$y$ and~$z$.
\[
  \begin{array}{l}
    \While{a = y}{\Skip} \seq {}\\
    \programfont{declassify}~z~\programfont{to}~\textsc{PublicTrusted}
  \end{array}
\]
A progress-insensitive definition should ignore the progress leak in the first line,
correctly identify that the attacker cannot influence the second line, and consider this program (progress-insensitively) robust.
However, an RD definition using a progress-insensitive notion of leakage for the first attack---the one used to check if leakage is allowed---%
would incorrectly classify the program.
Given secrets~$y_1$ and~$y_2$, the first attack could set~$a = y_1$,
sending $\trc_{11}$ into an infinite loop and causing all leakage in the first attack to stem from progress.
The result would satisfy any progress-insensitive notion of ``leaks nothing.''
However, a second attack where~$a$ differs from both~$y_1$ and~$y_2$
would cause~$\trc_{12}$ and~$\trc_{22}$ to both execute line two, producing more direct (non-progress) leakage and appear insecure.

Sending the program into an infinite loop causes the attacker to (voluntarily) learn \emph{less} than the developer intended,
and should be irrelevant to a progress-insensitive condition.
Prior work rules out such nontermination-based irrelevant attacks by demanding termination~\citep{MyersSZ06,ChongM06}.
The corresponding progress formulation would require the first attack to progress beyond the current execution point~$\pfx$ for both secrets.

Such a condition ensures the first attack does exhibit a progress leak up to~$\pfx$.
When combined with the existing PINI-style assumption,
Theorem~\ref{thm:psni-is-pini-lfp} tells us the attack cannot leak anything, regardless of the channel;
it must exhibit PSNI up to~$\pfx$.
We therefore require this condition directly and prohibit \emph{all} leakage between~$\trc_{11}$ and~$\trc_{21}$.
The definition of ``leaks nothing'' in the \emph{second} attack, however, determines the version of RD.

The second complication of considering leakage up to some prefix~$\pfx \leq \trc_{11}$
is that~$\pfx$ is only meant to limit the impact of \emph{endorsements}.
It is insufficient to simply require~$\trc_{21}$, the trace with the other secret, match~$p$ exactly.
Consider the following program with the same labels as above.
\[
  (\While{y}{\Skip}) \seq a \assign 5
\]
All leakage in this program is robust, as the attack has no impact on the behavior.
However, using secret values 0 and 1, if $\pfx \leq \trc_{11}$ happens to not include the assignment to~$a$---
though it exists in~$\trc_{11}$---$\trc_{21}$ could match it,
and the requirement suggested above would demand the second attack leak nothing, which is clearly false.

To handle this situation correctly, we formalize the intuition that the first attack satisfies PSNI up to~$\pfx$
by requiring it to hold for \emph{every} prefix of $\trc_{11}$ that is $\Trust$-equivalent to~$\pfx$.
Since~$a$ is untrusted, some $\pfx_{11} \leq \trc_{11}$ includes the assignment and satisfies $\pfx_{11} \lequiv[\Trust] \pfx$,
but~$\trc_{21}$ is stuck in an infinite loop, so it has no prefix $\Pub$-equivalent to~$\pfx_{11}$.
This more expansive premise recognizes the leakage in the first attack, thereby allowing the same leakage to occur in the second.

Taking this structure and using PSNI as the notion of ``leaks nothing'' for the second attack
gives us a complete definition of progress-sensitive robust declassification~(PSRD).
\[
  \def\arraystretch{1.2}
  \PsRd[(\Pub, \Trust)] \defeq \begin{array}[t]{@{}r@{~}c@{~}l@{}}
    \big\{ \T \subseteq \TT & \mid & \forall \trc_{11}, \trc_{12}, \trc_{21}, \trc_{22} \in \T\ldotp \nmifEqIn(\trc_{11}, \trc_{12}, \trc_{21}, \trc_{22}) \\
    && ~\implies
    \begin{array}[t]{@{}r@{\;}l@{}}
      \forall \pfx \leq \trc_{11}\ldotp &\big( \forall \pfx_{11} \leq \trc_{11}\ldotp \pfx_{11} \lequiv[\Trust] \pfx \implies \exists \pfx_{21} \leq \trc_{21}\ldotp \pfx_{11} \lequiv[\Pub] \pfx_{21} \big) \\
      \implies & \big( \forall \pfx_{12} \leq \trc_{12}\ldotp \pfx_{12} \lequiv[\Trust] \pfx \implies \exists \pfx_{22} \leq \trc_{22}\ldotp \pfx_{12} \lequiv[\Pub] \pfx_{22} \big) \big\}
    \end{array}
  \end{array}
\]
Here we use $\nmifEqIn$ to indicate that the initial states of all four traces properly correspond to pairs of attacks and secrets, defined by the following equivalences.
\[
  \nmifEqIn[(\Pub,\Trust)](\trc_{11}, \trc_{12}, \trc_{21}, \trc_{22}) \defeq
  \vcenter{\hbox{\begin{tikzpicture}[baseline=0]
    \node[inner sep=2pt] (t11) {$\Input(\trc_{11})$};
    \node[inner sep=2pt,right=2.5em of t11] (t21) {$\Input(\trc_{21})$};
    \node[inner sep=2pt,below=2em of t11] (t12) {$\Input(\trc_{12})$};
    \node[inner sep=2pt,below=2em of t21] (t22) {$\Input(\trc_{22})$};

    \path (t11) edge[draw=gray] node[inner sep=1pt,fill=white,yshift=-0.9pt] {$\lequiv[\Pub]$} (t21);
    \path (t12) edge[draw=gray] node[inner sep=1pt,fill=white,yshift=-0.9pt] {$\lequiv[\Pub]$} (t22);
    \path (t21) edge[draw=gray] node[inner sep=1pt,fill,color=white] (eq-trust1) {$\approx$} (t22);
    \path (t11) edge[draw=gray] node[inner sep=1pt,fill,color=white] (eq-trust2) {$\approx$} (t12);
    \node[yshift=-0.9pt,xshift=0.3pt] at (eq-trust1) {$\lequiv[\rlap{$\scriptstyle\Trust$}]$};
    \node[yshift=-0.9pt,xshift=0.3pt] at (eq-trust2) {$\lequiv[\rlap{$\scriptstyle\Trust$}]$};
  \end{tikzpicture}}}
\]

This definition follows exactly the intuition above.
Take any four traces whose inputs match as two pairs, with only secrets differing within a pair
and only untrusted (attacker) inputs differing between pairs.
For every endorsement-alignment prefix $\pfx \leq \trc_{11}$ of one trace,
if the first pair (attack) leaks nothing up to~$\pfx$, then neither does the second.

Like noninterference, this definition immediately extends to security against all attacks
by taking the intersection over the set~$\mathbb{A}$ of all attackers:
$\PsRd[] \defeq \bigcap_{\adv \in \mathbb{A}} \PsRd$.

Defining progress-insensitive robust declassification~(PIRD) requires only changing the definition of ``leaks nothing'' in the second attack.
Instead of prohibiting leakage through any channel up to~$\pfx$, PIRD allows leakage be due to progress.
We use the same approach as \PiNi[] and require the trace prefixes appear as prefixes of each other to a public observer.
\[
  \def\arraystretch{1.2}
  \PiRd[(\Pub, \Trust)] \defeq \begin{array}[t]{@{}r@{~}c@{~}l@{}}
    \big\{ \T \subseteq \TT & \mid & \forall \trc_{11}, \trc_{12}, \trc_{21}, \trc_{22} \in \T\ldotp \nmifEqIn(\trc_{11}, \trc_{12}, \trc_{21}, \trc_{22}) \\
    && ~\implies
    \begin{array}[t]{@{}r@{\;}l@{}}
      \forall \pfx \leq \trc_{11}\ldotp &\big( \forall \pfx_{11} \leq \trc_{11}\ldotp \pfx_{11} \lequiv[\Trust] \pfx \implies \exists \pfx_{21} \leq \trc_{21}\ldotp \pfx_{11} \lequiv[\Pub] \pfx_{21} \big) \\
      \implies & \big( \forall \pfx_{12} \leq \trc_{12}, \pfx_{22} \leq \trc_{22}\ldotp \pfx_{12} \lequiv[\Trust] \pfx \\
      & \quad \implies (\pfx_{12} \leq_\Pub \pfx_{22} \mathrel{\lor} \pfx_{12} \geq_\Pub \pfx_{22}) \big) \big\}
    \end{array}
  \end{array}
\]

By changing the notion of leakage in the second attack instead to prohibit leakage through progress, but allow leakage through event contents,
we acquire a robust declassification analogue of LFP we call \emph{robust progress leakage}~(RPL).
\[
  \def\arraystretch{1.2}
  \Rpl[(\Pub, \Trust)] \defeq \begin{array}[t]{@{}r@{~}c@{~}l@{}}
    \big\{ \T \subseteq \TT & \mid & \forall \trc_{11}, \trc_{12}, \trc_{21}, \trc_{22} \in \T\ldotp \nmifEqIn(\trc_{11}, \trc_{12}, \trc_{21}, \trc_{22}) \\
    && ~\implies
    \begin{array}[t]{@{}r@{\;}l@{}}
      \forall \pfx \leq \trc_{11}\ldotp &\big( \forall \pfx_{11} \leq \trc_{11}\ldotp \pfx_{11} \lequiv[\Trust] \pfx \implies \exists \pfx_{21} \leq \trc_{21}\ldotp \pfx_{11} \lequiv[\Pub] \pfx_{21} \big) \\
      \implies & \big( \forall \pfx_{12} \leq \trc_{12}, \pfx_{22} \leq \trc_{22}\ldotp \pfx_{12} \lequiv[\Trust] \pfx \\
      & \quad \implies \pfx_{22} <_\Pub \pfx_{12} \implies \Prog[\Pub](\pfx_{22}, \trc_{22}) \big) \big\}
    \end{array}
  \end{array}
\]

As with noninterference, enforcing PSRD is equivalent to enforcing both PIRD and RPL.
\begin{theorem}
  \label{thm:rd-split}
  For any attacker~$\adv$, $\PsRd = \PiRd \cap \Rpl$.
\end{theorem}

\paragraph{Hypersafety and Hyperliveness}
\label{sec:nmif-safety-and-liveness}

Much like \PsNi[], \PsRd[] is subset-closed, but neither hypersafety nor hyperliveness.
Unlike noninterference, however, the progress insensitivity and progress leakage split is not a decomposition into hypersafety and hyperliveness;
\PiRd[] is not hypersafety.

Formally, a hyperproperty~$H$ is hypersafety if, for any trace set~$\T$ violating~$H$ ($T \notin H$),
there is a finite set~$\{\pfx_i\}$ of finite prefixes such that
(1)~for each $\pfx_i$, there is some $\trc_i \in \T$ such that $\pfx_i \leq \trc_i$,
and (2)~if $\T'$ satisfies property~1, then $\T' \notin H$~\citep{hyperproperties10}.
$\PiRd$ does not satisfy this requirement.
Consider a set~$\T^\star$ of four traces, $\trc_{11}$, $\trc_{12}$, $\trc_{21}$, and $\trc_{22}$ where the trace inputs match as required by $\PiRd$,
$\trc_{11}$~and~$\trc_{21}$ are infinite but contain no visible events at all,
and $\trc_{12}$ and $\trc_{22}$ have different public-untrusted events---no traces contain trusted events.
Here $\trc_{11}$ and $\trc_{21}$ leak no information relative to each other, even through progress,
but $\trc_{12}$ and $\trc_{22}$ do, meaning $\T^\star \notin \PiRd$.
However, any finite prefix of $\trc_{11}$ can be extended with some public-untrusted event,
rendering the PSNI-style premise false, creating a trace set that \emph{does} satisfy $\PiRd$.
This counterexample is not verified in Rocq.

We leave decomposing these more complicated hyperproperties into hypersafety and hyperliveness as future work.

\subsection{Nonmalleable Information Flow}

While RD constrains declassification based on integrity,
recall that transparent endorsement~(TE) constrains endorsement based on confidentiality
and prohibits endorsement of secret information.
As in the original formulation~\citep{nmifc17}, TE is precisely dual to RD, switching the roles of the attacks and secrets.

This duality allows for immediate definitions of \PsTe, \PiTe, and \emph{transparent progress control}, \Tpc,
a prohibition on secrets influencing an attacker's control over progress.
\[
  \def\arraystretch{1.2}
  \Tpc[(\Pub, \Trust)] \defeq \begin{array}[t]{@{}r@{~}c@{~}l@{}}
    \big\{ \T \subseteq \TT & \mid & \forall \trc_{11}, \trc_{12}, \trc_{21}, \trc_{22} \in \T\ldotp \nmifEqIn(\trc_{11}, \trc_{12}, \trc_{21}, \trc_{22}) \\
    && ~\implies
    \begin{array}[t]{@{}r@{\;}l@{}}
      \forall \pfx \leq \trc_{11}\ldotp &\big( \forall \pfx_{11} \leq \trc_{11}\ldotp \pfx_{11} \lequiv[\Pub] \pfx \implies \exists \pfx_{12} \leq \trc_{12}\ldotp \pfx_{11} \lequiv[\Trust] \pfx_{12} \big) \\
      \implies & \big( \forall \pfx_{12} \leq \trc_{12}, \pfx_{22} \leq \trc_{22}\ldotp \pfx_{12} \lequiv[\Pub] \pfx \\
      & \quad \implies \pfx_{22} <_\Trust \pfx_{12} \implies \Prog[\Trust](\pfx_{22}, \trc_{22}) \big) \big\}
    \end{array}
  \end{array}
\]

Similarly, nonmalleable information flow is the intersection of RD and TE,
immediately giving progress-sensitive and insensitive definitions,
and a definition of \emph{nonmalleable progress leakage}, \NmPl, the progress-only counterpart to NMIF.
\begin{mathpar}
  \PsNmif \defeq \PsRd \cap \PsTe
  \and
  \PiNmif \defeq \PiRd \cap \PiTe
  \and
  \NmPl \defeq \Rpl \cap \Tpc
\end{mathpar}

%% file: calculus.tex
\section{A Core Calculus for Secure Progress Leakage}
\label{sec:calculus}

We now describe a core calculus that securely constrains progress leakage without eliminating it.
The calculus is a simple imperative calculus with only numeric values.
The syntax below contains the standard \Imp commands plus \PDown and \Stop.
\[
  \begin{array}{rcl}
    e & \Coloneqq & n \alt x \alt e \otimes e \\[0.5ex]
    c & \Coloneqq & \Skip \alt x \assign e \alt c \seq c \alt \ITE{e}{c}{c} \alt \While{e}{c} \alt \PDown_\ell c \alt \Stop
  \end{array}
\]

The \Stop command signals whole program termination.
It differs from \Skip, which indicates only that a given operation has no more steps.
Notably, \Stop should appear only on its own, does not appear in the surface language, and can never be well-typed (see Section~\ref{sec:type-system}).

The \PDown, or \emph{progress-downgrade}, operation is the main addition to the language.
It downgrades (declassifies and endorses) control flow, and thus termination behavior.
The command $\PDown_\ell~c$ means: run~$c$, then explicitly declassify (or endorse) the termination behavior of~$c$ to label~$\ell$.

To focus on progress leakage, we keep the calculus simple and omit data declassification and endorsement instructions.
Adding them with typing and semantic rules similar to \PDown would be straightforward and,
while it would likely not hamper security, the proofs would become considerably more involved.

\subsection{Operational Semantics}
\label{sec:semantics}

The semantics of the core calculus is mostly standard.
Expressions, which always terminate, use a big-step semantics,
and commands, which may diverge, use a small-step semantics.
A semantic configuration $\ctx{c}{\sigma}$ consists of a pair of a command~$c$ (or expression~$e$) and a memory~$\sigma$,
where $\sigma : \Vars \rightharpoonup \NN$ is a partial function from variable names to values (natural numbers).

\begin{figure}
  \begin{ruleset}
    \EStopRule \and \EAssignRule
    \\
    \ESeqStepRule \and \ESeqSkipRule
    \and
    \EPDownStepRule \and \EPDownSkipRule
  \end{ruleset}
  \caption{Selected operational semantic rules}
  \label{fig:semantics}
\end{figure}

Figure~\ref{fig:semantics} presents selected small-step operational semantics.
The complete semantics can be found in Appendix~\ref{app:full-defs}.
Note that both \ruleref{E-SeqStep} and \ruleref{E-PDownStep} require that the inductive step not produce~\Stop,
ensuring that \Stop appears only for full program termination.
Instead, \ruleref{E-SeqSkip} and \ruleref{E-PDownSkip} proceed directly when there is a \Skip.

Each semantic step also produces an event that will form the elements of an execution trace.
All but three steps produce~$\bullet$, indicating no effects or information.
\ruleref{E-Stop} produces~$\stopEff$, indicating that the program has terminated.
\ruleref{E-Assign} produces an assignment event,~$\assignEff(x, n)$,
indicating that variable~$x$ has been assigned value~$n$.
Finally, \ruleref{E-PDownSkip} produces a progress downgrade event,~$\termEff(\ell)$,
explicitly making visible to label~$\ell$ that the program has reached this point in execution,
and therefore the body of the $\PDown_\ell$ statement terminated.

\subsection{Type System}
\label{sec:type-system}

The type system has different judgments for expressions and commands.
Since all values are numeric, the types are simply information-flow labels.
Expression judgments take the form $\Gamma \proves e : \ell$,
where $\Gamma$ is a partial map from variable names to labels, and $\ell$ is a label,
and mean that label~$\ell$ is at least as restrictive as the security policy of any input to expression~$e$.
That is, it is safe to treat~$e$ as having policy~$\ell$.

\begin{figure}
  \centering
  \begin{mathpar}
    \SetRuleLabelLoc{left}
    \TVarianceRule \and \TSkipRule \and \TAssignRule
    \and
    \TIfRule \and \TSeqRule \and \TWhileRule
    \and
    \TPDownRule
  \end{mathpar}
  \caption{Typing rules for expressions and commands}
  \label{fig:typing}
\end{figure}

Command judgments take the form $\Gamma; \pc \proves c \tpout \nt$,
where $\Gamma$~is as before, and~$\pc$ and~$\nt$ are labels used to constrain effects.
The~$\pc$, or \emph{program counter}, label is standard~\citep{SabelfeldM03,HirschC21},
and it serves as both a lower bound on the visibility of a command's effects
and a means of controlling implicit information flows.
Flows can be either \emph{explicit}, when resulting from a direct assignment like $x \assign y$,
or \emph{implicit} when stemming from control flow.
Consider the following program.
$$\ITE{y}{x \assign 0}{x \assign 1}$$
This program only directly assigns constant values to~$x$, but the value of~$y$ implicitly influences~$x$.
The~$\pc$ label constrains these flows through two requirements:
(1)~the effects of an \IfN statement must be no more public than the condition,
and (2)~an assignment is an effect, meaning the~$\pc$ must flow to the label of the variable being assigned.
Type-checking the above example then requires $\Gamma(y) \flowsto \pc$ and $\pc \flowsto \Gamma(x)$,
transitivity ensuring $\Gamma(y) \flowsto \Gamma(x)$ and preventing information leakage.

The~$\nt$, or \emph{nontermination}, label constraints progress leakage.
It serves as an upper bound on the sensitivity of termination behavior of~$c$.
That is, anyone at or above~$\nt$ may safely learn if~$c$ terminates without leaking information.

Figure~\ref{fig:typing} presents the full type system.
\ruleref{Variance} formalizes the notion that $\pc$~is a \emph{lower} bound while $\nt$~is an \emph{upper} bound.
If~$c$ only produces high effects and its termination behavior is only influenced by low data,
it is safe to treat it as if it may produce lower effects---$\pc \flowsto \pc'$---and has termination influenced by higher data---$\nt' \flowsto \nt$.
Including this rule allows \ruleref{Assign}, \ruleref{If}, and \ruleref{While} to demand equality of labels, rather than flows,
simplifying the presentation and analysis.

\ruleref{Skip} and \ruleref{Assign} leave~$\nt$ unconstrained, as \Skip and assignments always terminate.
\ruleref{Assign} constrains explicit flows by requiring $\Gamma \proves e : \ell$ where $\ell$~is the label of the assigned variable.
Since assignments are effects, it also sets the~$\pc$ bound at~$\ell$.
\ruleref{If} requires $\Gamma \proves e : \pc$,
meaning the effects of the branches must be bounded below by the label of the condition, completing the implicit flow restriction.
As \IfN's termination behavior is that of its branches, $\nt$~remains unchanged.

The other rules more directly constrain termination leakage.
When sequentially composing commands~$c_1 \seq c_2$, if~$c_1$ diverges, then~$c_2$ will never execute,
meaning visible effects of~$c_2$ can leak whether or not~$c_1$ terminated.
To maintain security, \ruleref{Seq} therefore requires the effects of~$c_2$ to be bounded below by~$\nt_1$.
To ensure $\pc_1$~and~$\nt_2$ correctly bound the composed command,
$\pc_1$~must be a lower bound on the effects of both~$c_1$ and~$c_2$---$\Gamma; \pc_1 \proves c_1 \tpout \nt_1$ and $\pc_1 \flowsto \pc_2$, respectively.
Similarly, $\nt_2$~must be an upper bound on their termination sensitivity---$\nt_1 \flowsto \nt_2$ and $\Gamma; \pc_2 \proves c_2 \tpout \nt_2$, respectively.

\ruleref{While} uses the same label for the condition, the~$\pc$, and nontermination labels for three reasons.
First, the premise \mbox{$\Gamma \proves e : \pc$} ensures that the guard can safely influence effects in the loop.
Second, requiring~$c$ to type-check with a nontermination label of~$\pc$
ensures that learning if~$c$ terminates does not leak more information than~$c$'s own effects.
This requirement is necessary because the termination behavior of one execution of~$c$ can influence whether or not~$c$ executes again.
Third, both the loop condition and the termination behavior of~$c$ can influence if the entire loop terminates.
Since both have label~$\pc$, the nontermination label of the entire loop is~$\pc$.

Lastly, \ruleref{PDown} concerns explicit downgrades of progress information.
Recall that \ruleref{E-PDownSkip}, when executed, directly reveals to label~$\ell$ that the body of the $\PDown_\ell$ statement terminated.
As a result, the termination behavior of~$\PDown_\ell c$ leaks no \emph{additional} information to anyone at or above~$\ell$,
so \ruleref{PDown} sets the new nontermination label to~$\ell$.

Since $\PDown_\ell$ creates a visible effect at~$\ell$, the~$\pc$ must properly bound it: $\pc \flowsto \ell$.
Despite having a variance rule, this premise is a flow rather than an equality.
Downgrading to~$\pc$ and varying the nontermination label to~$\ell$ is semantically different from downgrading directly to~$\ell$.
The former releases information to~$\pc$, which may be lower than~$\ell$.

Enforcing NMIF requires restricting influence over both the content and timing of downgrades.
For a \PDown instructions, the old nontermination label~$\nt$ bounds what information might be released, so we require it be non-compromised.
Restricting the timing would normally involve similarly checking the~$\pc$ to avoid improper influence over the control flow.
However, if $\Gamma; \pc \proves c \tpout \nt$, then either $\pc \flowsto \nt$ or $\Gamma; \pc \proves c \tpout \ell'$ for any~$\ell'$.
In the first case, the antimonotonicity of~$\reflect$ together with~$\nt$ being non-compromised ensure $\pc \flowsto \reflect(\pc)$,
making such a premise redundant.
In the second case, setting $\ell' = \ell$ shows that there is no actual leakage to constrain.

\paragraph{Type Soundness}
\label{sec:lang-props}

This type system satisfies the basic soundness property that well-typed programs do not get stuck.
Since the type system presumes a mapping~$\Gamma$ of variables names to labels
and the semantics operates over a memory~$\sigma$, we require that all names referenced in~$\Gamma$ exist in~$\sigma$.

\begin{theorem}[Type Soundness]
  If $\Gamma; \pc \proves c \tpout \nt$, then for any memory~$\sigma$ where $\dom(\Gamma) \subseteq \dom(\sigma)$,
  if $\ctx{c}{\sigma} \stepsmany{} \ctx{c'}{\sigma'}$,
  then either $c' = \Stop$ or $\ctx{c'}{\sigma'}$ can step.
\end{theorem}

\subsection{Example Revisited}
\label{sec:example-revisited}

To see the use of these typing rules, and in particular \PDown,
we look at the attraction mapping example from Section~\ref{sec:intro}
using simple public/secret and trusted/untrusted policies.
We assumed that lines~\ref{ex:lst:loc} and~\ref{ex:lst:region} could both hang due to signal failures, revealing precise location.
The nontermination label~$\nt_\mathit{loc}$ of both lines would be secret,
but it would remain trusted, as the attacker cannot influence the user's location or the control flow to this point.
Line~\ref{ex:lst:map} will contact the server---a publicly-visible effect---requiring the $\pc$~label to be public.

With these labels, \ruleref{Seq} rule would identify the potential progress leakage and require an explicit declassification.
Since $\nt_\mathit{loc}$ is secret--trusted, it is not compromised,
so \ruleref{PDown} allows wrapping lines~\ref{ex:lst:loc} and~\ref{ex:lst:region} in \PDown to public--trusted.

The code for line~\ref{ex:lst:map} comes from an untrusted source.
To prevent it from modifying trusted data, it must type-check with an untrusted~$\pc$.
Importantly, if it contains a loop depending on secrets---opening the attack we aim to prevent---
\ruleref{While} requires the $\nt$~label to be secret--untrusted---\emph{a compromised label}.
\ruleref{PDown} does not allow downgrading this compromised control flow,
so \ruleref{Seq} ensures no later operation can create public or trusted effects.
Embedded in a larger system, this would inevitably fail to compile, identifying the dangerous termination channel.
Requiring line~\ref{ex:lst:map} to type-check with a \emph{non-compromised} $\nt$~label
prevents attacker-controlled termination leakage,
but allows a benign loop over attractions to place nearby ones on a map, which has a public--untrusted~$\nt$~label.

\subsection{Program Behavior and Indistinguishability}
\label{sec:behav-lequiv}

To prove the security of this calculus, we must first define traces and low-equivalence in this setting.
A trace consists of the program inputs, which we consider to be the initial memory~$\sigma$, and a possibly-infinite stream of events~$\st$.
Command~$c$ \emph{produces} a trace $\trace{\sigma}{\st}$, denoted $c \leadsto \trace{\sigma}{\st}$,
if~$c$ outputs precisely~$\st$ when run with initial memory~$\sigma$.
That is, if~$\ctx{c}{\sigma}$ terminates, then~$\st$ is the full list of events it produces, ending with $\stopEff$.
If~$\ctx{c}{\sigma}$ diverges, then~$\st$ is infinite and contains all events emitted during the execution.

The set of all traces~$c$ can produce is its \emph{behavior}:
\[
  \Behav(c) \defeq \{ \trc \mid c \leadsto \trc \}
\]

Recall from Section~\ref{sec:security} that our security hyperproperties
assume an indistinguishability relation~$\lequiv$ on finite trace prefixes
that is parameterized by a downward-closed label set~$\D$ representing low-sensitivity policies.
A finite trace prefix consists of an input---the initial memory---and a finite list of events,
so we need low-equivalence relations for each.

Two memories are $\D$-equivalent if they contain the same values for locations with labels in~$\D$, though they may differ elsewhere.
Formalizing this idea requires labels for memory locations,
so we parameterize the equivalence on both~$\D$ and a context~$\Gamma$ mapping locations to labels, as in the type system.
$\D$-equivalence, denoted $\sigma_1 \Mequiv \sigma_2$, then demands only that $\sigma_1$ and $\sigma_2$ agree on~$x$ when $\Gamma(x) \in \D$.
Formally,
\[
  \sigma_1 \Mequiv \sigma_2 \defIff \forall x\ldotp \Gamma(x) \in \D \implies \sigma_1(x) = \sigma_2(x).
\]

Two finite sequences of events are $\D$-equivalent if, at~$\D$, they appear to contain the same events in the same order.
We model a $\D$-observer as being unable to gain information from internal steps, signified by~$\bullet$ events,
and both assignments and progress downgrades at labels not in~$\D$.
To make this intuition precise, we define a silent-at-$\D$ predicate $\Silent(\alpha)$ as follows.
\begin{mathpar}[\rulefiguresize]
  \infer{ }{\Silent(\bullet)}
  \and
  \infer{\Gamma(x) \notin \D}{\Silent(\assignEff(x, n))}
  \and
  \infer{\ell \notin \D}{\Silent(\termEff(\ell))}
\end{mathpar}
Note two important design decisions.
First, the termination event~$\stopEff$ is always visible, regardless of the label,
formalizing that a progress-sensitive observer can always distinguish termination from silent infinite loops.
Second, all assignments to low memory locations are visible,
providing a strong security guarantee by modeling a powerful attacker that can continuously monitor public areas in memory.
Modeling explicit output is possible using distinguished memory addresses
and making more events silent, which cannot leak more information.

Equivalence of finite event sequences~$\evtseq_1$ and~$\evtseq_2$, denoted $\evtseq_1 \Tequiv \evtseq_2$,
then simply ignores silent events and requires the rest to be identical.
\begin{mathpar}[\rulefiguresize]
  \infer{ }{\e \Tequiv \e}
  \and
  \infer{
    \evtseq_1 \Tequiv \evtseq_2
  }{\alpha \cdot \evtseq_1 \Tequiv \alpha \cdot \evtseq_2}
  \and
  \infer{
    \evtseq_1 \Tequiv \evtseq_2 \\\\
    \Silent(\alpha)
  }{\alpha \cdot \evtseq_1 \Tequiv \evtseq_2}
  \and
  \infer{
    \evtseq_1 \Tequiv \evtseq_2 \\\\
    \Silent(\alpha)
  }{\evtseq_1 \Tequiv \alpha \cdot \evtseq_2}
\end{mathpar}

We can now define equivalence of trace prefixes, also parameterized by~$\Gamma$ and suggestively denoted~$\lequivG$,
by requiring the initial memories and event sequences both be equivalent.
That is, $\trace{\sigma_1}{\evtseq_1} \lequivG \trace{\sigma_2}{\evtseq_2} \defIff \sigma_1 \Mequiv \sigma_2 ~\text{and}~ \evtseq_1 \Tequiv \evtseq_2$.

Using these definitions of traces, prefixes, and equivalences is sufficient to state and prove the security of our core calculus.
Since our equivalences are all parameterized on~$\Gamma$, we will write $\PsNiG[]$, and similarly for other hyperproperties,
to indicate that we are using $\lequivG[-]$ as the equivalence relation.

\subsection{Proving Security}
\label{sec:enforcing-security}

While this calculus and its type system are designed to enforce progress-sensitive NMIF,
which allows for controlled data leakage that violates noninterference,
it remains useful to confirm that explicit downgrades are the \emph{only} way to violate noninterference.
The omission of data downgrade operations means all leakage should be through progress channels.
We verify this by proving that well-typed commands enforce PINI, which does not consider progress leakage.

\begin{theorem}[Progress-insensitive NI]
  \label{thm:pini}
  If $\Gamma; \pc \proves c \tpout \nt$, then $\Behav(c) \in \PiNiG[]$.
\end{theorem}

While some progress leakage is allowed, it should only come through two channels explicit in the type system:
\PDown instructions and programs with high~$\nt$ labels.
To eliminate leakage that the type system tracks and reports in the~$\nt$ labels, we can demand the command type-check with a low~$\nt$ label.
To ensure that all remaining leakage arises from \PDown instructions,
we define a notion of \emph{downgrade freedom} from a downward-closed set~$\D$.
Intuitively, a command that does not downgrade from outside~$\D$---a ``high'' label---to inside~$\D$---a ``low'' label---%
should also enforce LFP, and thus PSNI, at~$\D$.
We formalize this lack of downgrading as follows.

\begin{definition}[$\D$-downgrade Freedom]
  The proof $\Gamma; \pc \proves c \tpout \nt$ is \emph{$\D$-downgrade free} if,
  for every subcommand $\PDown_\ell c'$, either $\ell \notin \D$ or
  the subproof of $\Gamma; \pc \proves c' \tpout \nt'$ has $\nt' \in \D$.
\end{definition}

$\D$-downgrade freedom and a low~$\nt$ label are together sufficient to prevent progress leakage,
proving that all leakage is properly accounted for.
Together with Theorem~\ref{thm:pini}, this guarantees PSNI.
\begin{theorem}[$\D$-downgrade-free PSNI]
  \label{thm:down-free-psni}
  For any downward-closed label set~$\D$,
  if $\Gamma; \pc \proves c \tpout \nt$ with $\nt \in \D$ and the typing proof is $\D$-downgrade free, then $\Behav(c) \in \PsNiG$.
\end{theorem}

\paragraph{Enforcing Nonmalleable Information Flow}
For the same reason that Theorem~\ref{thm:down-free-psni} requires $\nt \in \D$,
a compromised~$\nt$ label---which cannot be safely downgraded---%
signals potentially insecure progress leakage.
As a result, not every well-typed command enforces nonmalleable progress leakage against every attacker.
However, if $\nt \in \Pub \cup \Trust$ for an attacker $\adv = (\Pub, \Trust)$,
then secret influence has been safely declassified, attacker influence has been safely endorsed, or both.
In each case, progress leakage will be robust, guaranteeing security.

\begin{theorem}[Low-$\nt$ Nonmalleable Progress Leakage]
  \label{thm:low-nt-nmpl}
  Given an attacker $\adv = (\Pub, \Trust)$, if $\Gamma; \pc \proves c \tpout \nt$ with $\nt \in \Pub \cup \Trust$, then $\Behav(c) \in \NmPlG$.
\end{theorem}

Also recall that non-compromised labels are public, trusted, or both for \emph{every} attacker (Proposition~\ref{prop:non-comp-low}).
Combined with Theorem~\ref{thm:low-nt-nmpl}, this provides the condition necessary to ensure security against all attackers.

\begin{theorem}[NMPL]
  \label{thm:nmpl}
  If $\Gamma; \pc \proves c \tpout \nt$ with $\nt \flowsto \reflect(\nt)$, then $\Behav(c) \in \NmPlG[]$.
\end{theorem}

Because noninterference (in particular \PiNi[]) is strictly stronger than the corresponding form of NMIF (\PiNmif[]),
Theorems~\ref{thm:rd-split},~\ref{thm:pini}, and~\ref{thm:nmpl} combine to show that all well-typed programs enforce progress-sensitive NMIF.

\begin{corollary}[Progress-Sensitive NMIF]
  \label{cor:psnmif}
  If $\Gamma; \pc \proves c \tpout \nt$ with $\nt \flowsto \reflect(\nt)$, then $\Behav(c) \in \PsNmifG[]$.
\end{corollary}

%% file: inference.tex
\section{Inferring Progress Downgrades}
\label{sec:inference}

One major advantage of enforcing security with a static type system is support for inference procedures.
To reduce programmer burden, they can omit explicit progress downgrades and instead
the compiler can infer their locations if any secure placement is possible.
Downgrades are generally considered sensitive operations requiring audits,
but a constructive inference procedure can direct the programmer to specific code points, minimizing manual effort.
We now present such an algorithm that is sound and complete with respect to the type system,
highly efficient, and infers a minimal set of downgrades.

\subsection{Label Structure}
\label{sec:inf-label-model}

The extreme generality of the label model in Section~\ref{sec:label-model} is good for expressive power,
but its lack of structure makes using it for computations challenging.
To make inference tractable, we require somewhat more structure on the labels.

First, flows-to ($\flowsto$) must to be antisymmetric in addition to reflexive and transitive.
There must be binary meet~($\meet$) and join~($\join$) functions
that compute the greatest lower bound and least upper bound of two labels, respectively.
These changes make~$\Labs$ a lattice.
Second, along with the global least label~$\bot$, we require a global greatest label~$\top$ where $\ell \flowsto \top$ for all $\ell \in \Labs$.

\subsection{Inference Algorithm}
\label{sec:inference-algo}

The inference algorithm, $\InferN$, consists of three parts.
The first, $\GetLabelN$, computes the minimum label of an expression~$e$ or determines that it cannot be typed.
The second, $\InfRelN$, does most of the work.
It places the progress downgrades, or determines that no placement will generate well-typed code.
It also records auxiliary information that the third part, $\RelSetN$, uses to set the label on downgrades placed by $\InfRelN$.

The goal is to infer a \emph{secure} placement of $\PDown$ instructions, slightly different than producing any well-typed command.
The type system guarantees progress-sensitive security only when the $\nt$~label is non-compromised (see Section~\ref{sec:enforcing-security}),
so $\InferN$ places $\PDown$ statements such that the resulting command is well-typed \emph{with a non-compromised nontermination label}.
Since $\InferN$ is complete with respect to the type system (Theorem~\ref{thm:complete-inf} below),
if any such placement exists, it will find one.
Otherwise, the type system cannot prove that all of the program's leakage is nonmalleable,
and $\InferN$ will fail, indicating this fact.

\paragraph{Expression Labels}
Computing labels of expressions is straightforward.
We parameterize $\GetLabelN$ on a typing context~$\Gamma$,
and specify it as a partial function that is defined precisely when~$e$ is well-typed.
It produces the most permissive (lowest) label consistent with the typing context.
It is defined as follows.
\[
  \begin{array}{rcl}
    \GetLabel(x) & = & \Gamma(x) \\
    \GetLabel(n) & = & \bot \\
    \GetLabel(e_1 \otimes e_2) & = & \GetLabel(e_1) \join \GetLabel(e_2)
  \end{array}
\]

\paragraph{Downgrade Placement}

\begin{figure}
  \centering
  \renewcommand{\arraystretch}{1.25}
  \SetAlgoVlined
  \begin{tabular}{r@{\hspace{2ex}}c@{\hspace{2ex}}l}
    $\InfRel*{\Skip}$ & $=$ & $(\partSkip, \top, \bot)$
    \\
    $\InfRel*{x \assign e}$ & $=$ &
    \parbox[t]{22.5em}{
      $\ell \from \GetLabel(e) \seq
        \Assert~\pc \join \ell \flowsto \Gamma(x) \seq
        (\partAssign{x}{e}, \Gamma(x), \bot)$
      }
    \\
    $\InfRel*{\ITE{e}{c_1}{c_2}}$ & $=$ &
    \parbox[t]{21em}{
      $\ell \from \GetLabel(e) \seq {}$ \\
      $(\partc[1], b_1, \nt_1) \from \InfRel*[\pc \join \ell]{c_1} \seq {}$ \\
      $(\partc[2], b_2, \nt_2) \from \InfRel*[\pc \join \ell]{c_2} \seq {}$ \\
      \eIf{$\nt_1 \join \nt_2 \flowsto \reflect(\nt_1 \join \nt_2)$}{
        $(\ITELab{e}{\ell}{\partc[1]}{\partc[2]}, b_1 \meet b_2, \nt_1 \join \nt_2)$
      }{
        $(\ITELab{e}{\ell}{(\PartPDown~\partc[1])}{\partc[2]}, b_1 \meet b_2, \nt_2)$
      }
    }
    \\
    $\InfRel*{c_1 \seq c_2}$ & $=$ &
    \parbox[t]{17em}{
      $(\partc[1], b_1, \nt_1) \from \InfRel*{c_1} \seq {}$ \\
      $(\partc[2], b_2, \nt_2) \from \InfRel*{c_2} \seq {}$ \\
      \eIf{$\nt_1 \flowsto b_2$}{
        $({\color{partCol}\partc[1] \seqlab{\nt_1} \partc[2]}, b_1 \meet b_2, \nt_1 \join \nt_2)$
      }{
        $({\color{partCol}(\PartPDown~\partc[1]) \seqlab{\bot} \partc[2]}, b_1 \meet b_2, \pc \join \nt_2)$
      }
    }
    \\
    $\InfRel*{\While{e}{c}}$ & $=$ &
    \parbox[t]{21.5em}{
      $\ell \from \GetLabel(e) \seq \Assert~\pc \join \ell \flowsto \reflect(\pc \join \ell) \seq {}$ \\
      $(\partc, b, \nt) \from \InfRel*[\pc \join \ell]{c} \seq {}$ \\
      \eIf{$\nt \flowsto b$}{
        $(\WhileLab{e}{\ell \join \nt}{\partc}, b \meet \reflect(\pc \join \ell), \nt \join \pc \join \ell)$
      }{
        $(\WhileLab{e}{\ell}{(\PartPDown~\partc)}, b \meet \reflect(\pc \join \ell), \pc \join \ell)$
      }
    }
  \end{tabular}
  \caption{Procedure for inferring types and \PDown placement}
  \label{fig:inference-procedure}
\end{figure}

The $\InfRelN$ algorithm determines both the placement of $\PDown$ instructions and the nontermination label~$\nt$ if inference is possible.
It is also parameterized on~$\Gamma$, and is a partial function from a~$\pc$ label and a command~$c$ (free from progress downgrades)
to a triple: a partial command~$\partc$, a bound label~$b$, and~$\nt$.
A \emph{partial command} is an intermediate data structure with the same structure as a command,
but without labels on $\PDown$ and with auxiliary label information for control structures---conditionals, sequences, and loops.
We write partial commands in \partColorText and with a tilde.

The bound label~$b$ indicates how much the~$\pc$ can rise before inference will fail.
That is, no $\PDown$ placement can allow~$c$ to type-check in context~$\Gamma$ with a non-compromised $\nt$~label
and a $\pc$~label above $\pc \join b$.
This label is important for efficiency in the sequence and \WhileN cases.
Placing a progress downgrade around the first command or the loop body, respectively,
can allow the second command or loop body to use a more permissive~$\pc$.
A naive algorithm would thus make two recursive calls, one for each value of~$\pc$, resulting in exponential run time.
The bound label allows us to replace the second recursive call with a single flow check, drastically improving efficiency.

Figure~\ref{fig:inference-procedure} presents the full $\InfRelN$ algorithm.
For \Skip, the bound label is~$\top$ and nontermination label of~$\bot$,
since \Skip type-checks with any~$\pc$ and~$\nt$.
The assignment case is only slightly more complex.
It asserts that $\pc \join \ell \flowsto \Gamma(x)$, as the command will never type-check if that flow does not hold.
If it does hold, $\InfRelN$ sets $b = \Gamma(x)$, as the~$\pc$ cannot rise above that level and still produce a valid typing proof,
and $\nt = \bot$, as the command always terminates.

For conditionals, $\InfRelN$ first determines the label~$\ell$ of the condition,
before making recursive calls on both branches, using $\pc \join \ell$ as the~$\pc$ label.
Inference fails if the condition is ill-typed or either recursive call fails.
If they all succeed, the only remaining check is whether a downgrade is required.

Without a downgrade, the nontermination label of the conditional will be $\nt_1 \join \nt_2$,
where~$\nt_1$ and~$\nt_2$ are the nontermination labels of the branches.
If that join is non-compromised, no downgrade is needed.
If the join \emph{is} compromised, a downgrade is required around one branch.
Because both~$\nt_i$ are outputs of $\InfRelN$, either $\nt_i = \bot$ or $\pc \join \ell \flowsto \nt_i$ and both are non-compromised.
That means whenever $\nt_1 \join \nt_2$ is compromised, neither label is~$\bot$, so $\pc \join \ell \flowsto \nt_i$.
Downgrading the progress of either branch to $\pc \join \ell$
will therefore result in the entire \IfN statement having a non-compromised $\nt$~label,
so we arbitrarily choose to downgrade the \ThenN branch.
Some programs may also need to downgrade the termination behavior of the full \IfN statement---and thus the \ElseN branch.
Inserting \PDown into only the \ThenN branch preserves the ability to safely perform such a larger downgrade
while optimistically hoping it will be unnecessary.

For the conditional's bound label, a higher~$\pc$ must flow to both~$b_1$ and~$b_2$ for inference to succeed,
which corresponds precisely to a bound of their meet $b_1 \meet b_2$.

Sequential composition is where the bound label becomes important.
\ruleref{Seq} requires the~$\pc$ of~$c_2$ to be at least as high both the~$\pc$ and~$\nt$ labels of~$c_1$.
In the absence of a downgrade, $c_2$~must therefore type-check with $\pc \join \nt_1$.
A naive strategy would make a recursive call on~$c_1$ and then try to infer downgrades for~$c_2$ with $\pc \join \nt_1$.
If this inference on~$c_2$ fails, however, it may still be possible by downgrading~$c_1$'s progress
and infer downgrades on~$c_2$ using~$\pc$, requiring a second recursive call.

The bound label allows us to avoid this double recursive call and the resulting exponential running time.
Instead, $\InfRelN$ makes one recursive call on each of~$c_1$ and~$c_2$ using~$\pc$ for both.
If the nontermination label~$\nt_1$ of~$c_1$ flows to the bound label~$b_2$ of~$c_2$,
inference will still succeed on~$c_2$ when run with $\pc \join \nt_1$, and no downgrade is needed.
If $\nt_1 \nflowsto b_2$, inference with $\pc \join \nt_1$ would fail and a progress downgrade is required.

Additionally, the partial command for sequence includes an auxiliary label.
This label represents the amount~$c_2$'s program counter must increase beyond~$\pc$ in the typing proof.
Without a downgrade, this value is~$\nt_1$.
With a downgrade, no increase is required, so~$\bot$ suffices.

The \WhileN case first computes the label~$\ell$ of the condition and asserts that $\pc \join \ell \flowsto \reflect(\pc \join \ell)$.
The nontermination label of the loop cannot be lower than $\pc \join \ell$,
so this check is needed to ensure $\InfRelN$ only outputs non-compromised nontermination labels.
If the check passes, then it makes a recursive call on~$c$ using $\pc \join \ell$.
Since \WhileN loops can sequence~$c$ with itself, we again need to check if the first command's~$\nt$ label flows to the second's bound.
As both commands are the same, this check becomes $\nt \flowsto b$.
As in the sequence case above, a downgrade is necessary if and only if the flow does not hold.

The bound label for \WhileN is the meet of the recursively computed bound~$b$ and $\reflect(\pc \join \ell)$.
The first part captures the bound for the subcommand~$c$.
The second part ensures that $\pc \join \ell$ will remain non-compromised even when raising the~$\pc$.

The nontermination label is either $\nt \join \pc \join \ell$ if there is no downgrade, as all three impact termination,
or $\pc \join \ell$ if a progress downgrade lowered the inner nontermination label.

Finally, we include an auxiliary label as in the sequence case.
With no downgrade, the~$\pc$ in the body must rise by $\ell \join \nt$, and with a downgrade, only~$\ell$.

\paragraph{Setting Downgrade Labels}

\begin{figure}
  \centering
  \renewcommand{\arraystretch}{1.25}
  \begin{tabular}{r@{\hspace{2ex}}c@{\hspace{2ex}}l}
    $\RelSet*{\partSkip}$ & $=$ & $\Skip$
    \\
    $\RelSet*{\partAssign{x}{e}}$ & $=$ & $x \assign e$
    \\
    $\RelSet*{\ITELab{e}{\ell}{\partc[1]}{\partc[2]}}$ & $=$ &
    $\ITE*{e}{(\RelSet*[\pc \join \ell]{\partc[1]})}{(\RelSet*[\pc \join \ell]{\partc[2]})}$
    \\
    $\RelSet*{\partc[1] \seqlab{\ell} \partc[2]}$ & $=$ &
    $\RelSet*{\partc[1]} \seq \RelSet*[\pc \join \ell]{\partc[2]}$
    \\
    $\RelSet*{\WhileLab{e}{\ell}{\partc}}$ & $=$ & $\While{e}{\RelSet*[\pc \join \ell]{\partc}}$
    \\
    $\RelSet*{\color{partCol}\PartPDown~\partc}$ & $=$ & $\PDown_\pc \RelSet*{\partc}$
  \end{tabular}
  \caption{Procedure for setting \PDown labels}
  \label{fig:pd-lab-set}
\end{figure}

For sequential composition, $\InfRelN$ cannot determine the~$\pc$ of~$c_2$ until after the recursive call completes (and similarly for loops).
That means it does not have enough information to set the labels on \PDown instructions as it places them.
Instead, it embeds auxiliary label information in its output,
which $\RelSetN$ (Figure~\ref{fig:pd-lab-set}) uses on a second pass to set those labels.
For each command, it recursively executes on each sub-command,
increasing the~$\pc$ by the label specified in the auxiliary information,
and sets the label of each \PDown command to the current~$\pc$ as it goes.

The full inference algorithm first runs $\InfRelN$ and then $\RelSetN$ on its output.
\[
  \Infer{c} = (\partc, b, \nt) \from \InfRel*{c} \seq (\RelSet*\partc, \nt)
\]

\subsection{Soundness, Completeness, and Correctness}

The inference algorithm is sound and complete with respect to the type system using only non-compromised nontermination labels.
Formally, soundness says that, if the algorithm succeeds, then the resulting command is well-typed with a non-compromised $\nt$~label.

\begin{theorem}[Sound Inference]
  If $\Infer{c} = (c', \nt)$, then $\Gamma; \pc \proves c' \tpout \nt$ and $\nt \flowsto \reflect(\nt)$.
\end{theorem}

Intuitively, completeness says that, if there is any way to place \PDown instructions such that the resulting command is well-typed
with a non-compromised $\nt$~label, then $\InferN$ will find one.
To formalize this intuition, we use a downgrade-erasure operation, denoted $\erase{c}$ and defined as follows.
\[
  \begin{array}{r@{\hspace{0.5em}}c@{\hspace{0.5em}}l}
    \erase{\Skip} & = & \Skip \\
    \erase{x \assign e} & = & x \assign e \\
    \erase{c_1 \seq c_2} & = & \erase{c_1} \seq \erase{c_2} \\
  \end{array}
  \qquad
  \begin{array}{r@{\hspace{0.5em}}c@{\hspace{0.5em}}l}
    \erase{\ITE{e}{c_1}{c_2}} & = & \ITE{e}{\erase{c_1}}{\erase{c_2}} \\
    \erase{\While{e}{c}} & = & \While{e}{\erase{c}} \\
    \erase{\PDown_\ell c} & = & \erase{c}
  \end{array}
\]
The formal definition of completeness is then as follows.
\begin{theorem}[Complete Inference]
  \label{thm:complete-inf}
  For any command~$c$ and label~$\nt$ such that $\Gamma; \pc \proves c \tpout \nt$ and $\nt \flowsto \reflect(\nt)$,
  there is some~$c'$ and~$\nt'$ such that $\Infer{\erase{c}} = (c', \nt')$.
\end{theorem}

Finally, $\InferN$ is \emph{correct} in that it does not modify commands aside from possibly adding downgrade instructions.
\begin{theorem}[Correct Inference]
  If $\Infer{c} = (c', \nt)$, then $c = \erase{c'}$.
\end{theorem}

\subsection{Efficiency and Minimality}

The inference algorithm also aims to be efficient and only insert downgrades where necessary.
Achieving either goals is simple:
inserting downgrades everywhere is efficient but not minimal,
while trying every combination of downgrades and selecting a minimal well-typed one is minimal but highly inefficient.
The $\InferN$ algorithm accomplishes both.

To remain efficient, $\InferN$ consists of two sequential linear passes: $\InfRelN$ and $\RelSetN$.
In each pass, the only operations that could impact performance are label operations (join, meet, reflection, and flows-to checks).

For simple label models, like small finite lattices, these operations are constant time, producing linear efficiency.
More complicated label models may result in more overhead, but the structure of the algorithm mitigates some concern.
All flows-to checks query if a nontermination label---always the join of (at most) linearly many labels---
flows to a bound label---always the meet of (at most) linearly many labels.
Common security lattices, including subset lattices of permissions~\citep{histar11},
and free distributive lattices over a set of principals~\citep{MyersL98,flam15,lio11}
can run these checks very efficiently.

\paragraph{Minimality}

Even nonmalleable downgrades represent possible points of data leakage or corruption, so $\InferN$ aims to insert a minimal set.
However, there are many ways to define ``minimal.''
Semantic minimality---executing as few downgrades as possible---is appealing,
but the semantically-minimal set of downgrades for a program might not be statically well-defined; it could depend on the inputs.

Instead, we achieve a local syntactic notion of minimality:
removing any downgrade inserted by $\InferN$ (without adding another elsewhere) will always be ill-typed.
We formalize the idea of ``removing a downgrade'' using a relation~$\pdle$ to denote that two commands have the same structure,
but one may have more syntactic downgrade instructions than the other and the $\PDown$ labels may not match.
We define~$\pdle$ as the smallest structurally compatible preorder on commands---
it is reflexive, transitive, and admits structurally recursive rules like
$\frac{
  c_1' \mkern3mu\pdle\mkern3mu c_1
  \quad
  c_2' \mkern3mu\pdle\mkern3mu c_2
}{c_1' \mkern2mu\seq\mkern2mu c_2' \mkern3mu\pdle\mkern3mu c_1 \mkern2mu\seq\mkern2mu c_2}$---admitting the following rules.
\begin{mathpar}[\rulefiguresize]
  \infer{c' \pdle c}{c' \pdle \PDown_\ell c}
  \and
  \infer{c' \pdle c}{\PDown_{\ell'} c' \pdle \PDown_\ell c}
\end{mathpar}
Letting $c \pdeq c'$ denote commands with identical structure---they are the same except for the labels on $\PDown$ instructions---
this relation allows us to formalize syntactic minimality.
\begin{theorem}[Minimal Inference]
  If $\Infer{c_{\rm in}} = (c, \nt)$, then for any command~$c'$ where $c' \pdle c$,
  if $\Gamma; \pc \proves c' \tpout \nt'$ with $\nt' \flowsto \reflect(\nt')$,
  then $c' \pdeq c$.
\end{theorem}

%% file: rocq-overview.tex
\section{Proof Approach and Rocq Details}
\label{sec:rocq}

All theorems in this paper are mechanically verified in The Rocq Prover~\citep{rocq}
and are available online~\citep{rocq-artifact}.
We encode expressions and commands with a deep embedding as inductive types
and use \option types to encode partial functions, including typing contexts, memories, and~$\InferN$.
We assume decidable equality for variable names,
decidable set inclusion for label sets and attackers in the security theorems,
and decidable flows-to in the inference algorithm.

The proofs use the most general label model presented, but we verify that the less-general model
that explicitly separates confidentiality and integrity is actually a special case (see Section~\ref{sec:general-labels}).
There are also minor differences in the definitions of $\D$-equivalence of events and stores to make proofs simpler,
so we include definition equivalence proofs for each.

\subsection{Proving Security}

The proofs of noninterference (Theorems~\ref{thm:pini} and~\ref{thm:down-free-psni}) and NMPL (Theorem~\ref{thm:nmpl})
use the bridge-step relation introduced by~\citet{BayA20}, which defines a configuration emitting a $\D$-visible event.
That is, $\ctx{c}{\sigma}$ takes one or more steps, where only the last one produces something visible to~$\D$.
Formally, bridge steps are defined by the following inductive relation.
\begin{mathpar}[\rulefiguresize]
  \infer{
    \ctx{c}{\sigma} \stepsone[\alpha] \ctx{c'}{\sigma'} \\
    \neg\Silent(\alpha)
  }{\ctx{c}{\sigma} \bstep{\alpha} \ctx{c'}{\sigma'}}
  \and
  \infer{
    \ctx{c}{\sigma} \stepsone[\alpha'] \ctx{c'}{\sigma'} \\
    \Silent(\alpha') \\\\
    \ctx{c'}{\sigma'} \bstep{\alpha} \ctx{c''}{\sigma''}
  }{\ctx{c}{\sigma} \bstep{\alpha} \ctx{c''}{\sigma''}}
\end{mathpar}
A critical lemma shows that running one command with two $\D$-equivalent memories
produces the same bridge step unless one configuration silently diverges, in which case we can bound the visible events and nontermination label.
\begin{lemma}[Matching Bridge Step]
  For any command~$c$ where $\Gamma; \pc \proves c \tpout \nt$ and memories $\sigma_1$~and~$\sigma_2$ where $\sigma_1 \Mequiv \sigma_2$,
  if $\ctx{c}{\sigma_1} \bstep{\alpha} \ctx{c'}{\sigma_1'}$,
  then either
  \begin{itemize}[leftmargin=*,nosep]
    \item $\ctx{c}{\sigma_2} \bstep{\alpha} \ctx{c'}{\sigma_2'}$ with $\sigma_1' \Mequiv \sigma_2'$, or
    \item $\ctx{c}{\sigma_2}$ silently diverges---diverges without ever producing a $\D$-visible event---%
      and either $\alpha = \termEff(\ell)$ or both $\alpha = \Stop$ and $\nt \notin \D$.
  \end{itemize}
\end{lemma}

This lemma relies on a standard containment lemma.
\begin{lemma}[Containment]
  Given any downward-closed set~$\D$,
  if $\Gamma; \pc \proves c \tpout \nt$ with $\pc \notin \D$ and $\ctx{c}{\sigma} \stepsone[\alpha] \ctx{c'}{\sigma'}$,
  then $\sigma \Mequiv \sigma'$ and either $\Silent(\alpha)$ or $\alpha = \Stop$.
\end{lemma}

\paragraph{Nontermination}
Since our theorems focus on distinguishing terminating executions from nonterminating ones, their proofs rely on a similar differentiation.
The proof of progress-sensitive NMPL (Theorem~\ref{thm:nmpl}) uses the following lemma.
\begin{lemma}[While Trilemma]
  \label{li:while-trilemma}
  If a \WhileN loop is never stuck, then either ({\it i})~it terminates,
  or ({\it ii})~after some finite number of iterations, the body diverges,
  or ({\it iii})~the body converges on every iteration, but the loop executes infinitely many times.
\end{lemma}
This lemma is not provable in a constructive logic like Rocq;
deciding which branch holds requires solving the halting problem.
Instead, we verify that it holds classically---ensuring logical consistency---and take it as an axiom for the theorems.

\paragraph{A note on computability theory}
Lemma~\ref{li:while-trilemma} is not just undecidable,
no recursive enumeration procedure can determine the cause of nontermination and separate cases~({\it ii}) and~({\it iii}).
One can, however, recursively enumerate the loops satisfying~({\it ii}) \emph{using a halting oracle}.
Deciding the while trilemma is thus under $\mathbf{0}''$~\citep[for background, see, e.g.,][]{Soare16}.
Indeed, the Rocq proof of Lemma~\ref{li:while-trilemma} uses the classical assumption exactly twice:
once to separate terminating loops from divergent ones, and a second time to differentiate the cause of nontermination.

\subsection{Inference Properties}

Recall from Section~\ref{sec:inference-algo} that, $\InfRelN$ returns a bound label~$b$ in addition to a command and nontermination label.
The soundness and minimality of $\InferN$ rely on~$b$ properly bounding how much the~$\pc$ can rise before inference fails.
The following lemma formalizes this requirement.
\begin{lemma}[Bound Validity]
  If $\InfRel*{c} = (c', b, \nt)$, then for any non-compromised label~$\ell$,
  $\ell \flowsto b$ if and only if $\InfRel*[\pc \join \ell]{c}$ is defined.
\end{lemma}

Minimality also requires that the inferred nontermination label~$\nt$ be the smallest possible label.
\begin{lemma}[Least $\nt$]
  If $\Infer{c_{\rm in}} = (c, \nt)$, then for any~$c'$ and~$\nt'$
  if $c' \pdeq c$ and $\Gamma; \pc \proves c' \tpout \nt'$, then $\nt \flowsto \nt'$.
\end{lemma}

%% file: related.tex
\section{Related Work}
\label{sec:related}

We now discuss prior work on progress-sensitive security, secure downgrading, and information-security hyperproperties.

\paragraph{Progress-Sensitive Security}

Early type-based enforcement of termination-sensitive noninterference
either limit loops to only execute in fully public environments ($\pc = \bot$) and have fully public conditions~\citep{VolpanoS97,ONeillCC06}
or operate in a pure $\lambda$-calculus with call-by-name semantics~\citep{dcc99}.
These constraints led most work to target progress-insensitive security.
However, \citet{AskarovHSS08} show how progress channels can leak arbitrary data in effectful languages,
identifying a major risk of using progress-insensitive security with active attackers.

\citet{MooreAC12} provide a more precise type system similar to the one in Section~\ref{sec:type-system}.
To further relax the type system's restrictions, they include a progress downgrade operation they call~$\mathit{cast}$.
Notably, the type system does not restrict $\mathit{cast}$.
Instead, there is one semantics that appeals to a halting oracle and gets stuck if $\mathit{cast}$ could leak anything,
and a second that tracks a quantitative leakage budget.

\citet{BayA20} show how to define progress leakage as declassification in a progress-sensitive context
and introduce $\mathsf{tini}$ blocks much like our $\PDown$.
They consider only confidentiality and bound declassifications by a separate notion of declassifier authority, with no ability to ensure robustness.

\paragraph{Secure Downgrading}

Prior work on secure downgrades in IFC systems is extensive.
Some allow labels to specify what downgrades directly~\citep{LiZ05,ChongM08,lifty20}.
Delimited release allows declassifications based on syntactic code structure~\citep{delimitedRelease03}.
Intransitive information flow restricts flows based on policies that may not be transitive~\cite{Pinsky95,RoscoeG99,MantelS04,vanderMeyden07}.
None of these ideas use confidentiality and integrity to constrain each other, and so cannot consider nonmalleability concerns.

The original formulation of robust declassification~(RD) appears progress-sensitive, but gives no suggestions for enforcement~\citep{ZdancewicM01}.
The first enforcement mechanisms limit declassifications based on integrity levels,
but both enforce only progress-insensitive definitions of RD~\citep{ChongM06,MyersSZ06}.
\citet{AskarovM11} use knowledge-based formulations to define both progress-sensitive and progress-insensitive forms of RD,
but the definitions and enforcement are tailored to a four-point lattice
and the only suggestion for enforcing progress-sensitive RD is to eliminate progress leakage entirely.
\mbox{\citet{McCallBJ23}} provide a knowledge-based definition of RD for use in the challenging setting of reactive web applications.
Their definition is progress-insensitive
and relies on trace events that track syntactic downgrades,
making it intensional and hard to generalize to languges that do not directly track leakage.

\citet{nmifc17} define transparent endorsement as the dual of RD,
combine the two into nonmalleable information flow, and present all three as hyperproperties.
Their definition relies on traces having matching public-trusted events, making it inherently progress-insensitive,
and they only enforce it in a fully terminating language.
\citet{SolovievBG24} reformulate RD and NMIF in modal logic using Kripke frames, including both progress-sensitive and progress-insensitive formulations,
but like the progress-sensitive RD definitions, there is no enforcement mechanism.

\paragraph{Information-Security Hyperproperties}

Information-security conditions have been key examples of hyperproperties since their introduction.
\citet{hyperproperties10} show how to express multiple versions of noninterference as hyperproperties.
One such notion is \emph{observational determinism}~(OD)~\citep{McLean92,Roscoe95}, which \citet{ZdancewicM03} formulate similarly to noninterference,
with explicit consideration of trace prefixes.
\citeauthor{hyperproperties10} present OD as a hyperproperty similar to our definition of $\PiNi$ in Section~\ref{sec:ni-lfp}.

A variety of tools aim to specify or verify security-oriented hyperproperties.
Relational Hoare Type Theory~(RHTT)~\citep{NanevskiBG11} allows for precise specification of 2-hypersafety properties like noninterference.
Cartesian Hoare Logic~(CHL)~\citep{SousaD16} generalizes RHTT to arbitrary $k$-hypersafety properties for relational traces (input/output pairs).
As our main hyperproperties require considering intermediate trace prefixes of four traces, neither RHTT nor CHL can represent them.

Other tools and techniques aim to verify various forms of hyperproperties~\citep{LamportS21,BeutnerF22,UnnoTK21,CoenenFST19,FarzanV19},
with entirely different goals from this work.
They aim to verify application-specific hyperproperties (within a certain class), sometimes at great computational expense.
This work identifies highly-general security hyperproperties and enforces them through inexpensive type checking.
I hope the active research into verification will complement the results in this paper.

%% file: conclusion.tex
\section{Conclusion}
\label{sec:conclusion}

Information-flow control systems have long faced a tension between
providing strong whole-program security guarantees
and supporting necessary programming constructs, like declassification and loops.
Noninterference is too restrictive, but uncontrolled downgrading complicates stating and proving security.
Similarly, progress-insensitive security can leak arbitrary data to attackers who can influence nontermination,
but requiring loops to condition only on public data makes many programs difficult to write.

To resolve this tension, we distilled the separation between progress-sensitivity
and progress-insensitivity into a new hyperproperty called \emph{leakage-free progress},
and generalized it as \emph{nonmalleable progress leakage}~(NMPL),
an adaptation of the intuitions of nonmalleable information flow~(NMIF) to secure progress channels.
We explored how to enforce NMPL with a simple information-flow type system,
and finally we showed how to efficiently infer the locations of necessary progress downgrade operations
to aid developers or verify NMPL without explicit annotations.

We hope these foundations will support more expressive and powerful security-typed languages in the future.
This work used an imperative core calculus without data downgrades
to better focus on the core contribution and simplify formalisms and proofs.
Extending these ideas to more practical languages would be valuable future work,
with higher-order stateful languages posing a particularly interesting challenge.

%% file: acks.tex
\section*{Acknowledgments}
\label{sec:acks}

This paper has only one author, but many people helped make it possible.
Mike Hicks provided direction about which questions would be most impactful.
Leo Lampropoulos suggested a suitable notion for minimality of the downgrade inference algorithm.
Andrew K. Hirsch provided valuable comments on the text and,
along with Tej Chajed, helped work through multiple challenges in the Rocq code.
Additional thanks to Andrew C. Myers and Ashley Samuelson for help editing.
Finally, the shepherd and other anonymous reviewers
provided valuable feedback and suggestions for improving presentation.

%% file: full-calculus.tex
\section{Full Calculus Rules}
\label{app:full-defs}

We now present the full semantic and typing rules of the core calculus from Section~\ref{sec:calculus}.
The big-step semantic rules for expression are in Figure~\ref{fig:expr-semantics}
and the small-step rules for all commands are in Figure~\ref{fig:full-cmd-semantics}.
The typing rules for expressions are in Figure~\ref{fig:expr-typing}.
Figure~\ref{fig:typing} in Section~\ref{sec:type-system} contains all typing rules for commands.

\begin{figure}[h]
  \rulefiguresize
  \flushleft
  \fbox{$\ctx{e}{\sigma} \Downarrow n$}
  \begin{ruleset}
    \infer{ }{\ctx{x}{\sigma} \Downarrow \sigma(x)}
    \and
    \infer{ }{\ctx{n}{\sigma} \Downarrow n}
    \and
    \infer{
      \ctx{e_1}{\sigma} \Downarrow n_1 \\
      \ctx{e_2}{\sigma} \Downarrow n_2
    }{\ctx{e_1 \otimes e_2}{\sigma} \Downarrow (n_1 \otimes n_2)}
  \end{ruleset}
  \caption{Big-step operational semantics for expressions.}
  \label{fig:expr-semantics}
\end{figure}
\begin{figure}[h]
  \rulefiguresize
  \flushleft
  \fbox{$\ctx{e}{\sigma} \stepsone[\alpha] \ctx{e'}{\sigma'}$}
  \begin{ruleset}
    \EStopRule \and \EAssignRule
    \\
    \ESeqStepRule \and \ESeqSkipRule
    \\
    \EIfNRule[left] \and \EIfZRule[left]
    \\
    \EWhileRule
    \and
    \EPDownStepRule[left] \and \EPDownSkipRule[left]
  \end{ruleset}
  \caption{Full small-step operational semantics for commands.}
  \label{fig:full-cmd-semantics}
\end{figure}

\begin{figure}
  \rulefiguresize
  \flushleft
  \fbox{$\Gamma \proves e : \ell$}
  \begin{ruleset}
    \infer{\Gamma(x) = \ell}{\Gamma \proves x : \ell}
    \and
    \infer{ }{\Gamma \proves n : \ell}
    \and
    \infer{
      \Gamma \proves e_1 : \ell \\\\
      \Gamma \proves e_2 : \ell
    }{\Gamma \proves e_1 \otimes e_2 : \ell}
    \and
    \infer{
      \Gamma \proves e : \ell' \\\\
      \ell' \flowsto \ell
    }{\Gamma \proves e : \ell}
    \vspace{0.5\baselineskip}
  \end{ruleset}
  \caption{Typing rules for expressions.}
  \label{fig:expr-typing}
\end{figure}

%% file: ms.bbl

\begin{thebibliography}{52}


\ifx \showCODEN    \undefined \def \showCODEN     #1{\unskip}     \fi
\ifx \showDOI      \undefined \def \showDOI       #1{#1}\fi
\ifx \showISBNx    \undefined \def \showISBNx     #1{\unskip}     \fi
\ifx \showISBNxiii \undefined \def \showISBNxiii  #1{\unskip}     \fi
\ifx \showISSN     \undefined \def \showISSN      #1{\unskip}     \fi
\ifx \showLCCN     \undefined \def \showLCCN      #1{\unskip}     \fi
\ifx \shownote     \undefined \def \shownote      #1{#1}          \fi
\ifx \showarticletitle \undefined \def \showarticletitle #1{#1}   \fi
\ifx \showURL      \undefined \def \showURL       {\relax}        \fi
\providecommand\bibfield[2]{#2}
\providecommand\bibinfo[2]{#2}
\providecommand\natexlab[1]{#1}
\providecommand\showeprint[2][]{arXiv:#2}

\bibitem[Abadi et~al\mbox{.}(1999)]%
        {dcc99}
\bibfield{author}{\bibinfo{person}{Mart{\'i}n Abadi}, \bibinfo{person}{Anindya
  Banerjee}, \bibinfo{person}{Nevin Heintze}, {and} \bibinfo{person}{Jon
  Riecke}.} \bibinfo{year}{1999}\natexlab{}.
\newblock \showarticletitle{A Core Calculus of Dependency}. In
  \bibinfo{booktitle}{\emph{26\textsuperscript{th} ACM SIGPLAN Symposium on
  Principles of Programming Languages (POPL~'99)}}.
\newblock
\urldef\tempurl%
\url{https://doi.org/10.1145/292540.292555}
\showDOI{\tempurl}


\bibitem[Abate et~al\mbox{.}(2019)]%
        {journey-beyond19}
\bibfield{author}{\bibinfo{person}{Carmine Abate}, \bibinfo{person}{Roberto
  Blanco}, \bibinfo{person}{Deepak Garg}, \bibinfo{person}{C\u{a}t\u{a}lin
  Hri\c{t}cu}, \bibinfo{person}{Marco Patrignani}, {and}
  \bibinfo{person}{J\'{e}r\'{e}my Thibault}.} \bibinfo{year}{2019}\natexlab{}.
\newblock \showarticletitle{Journey Beyond Full Abstraction: Exploring Robust
  Property Preservation for Secure Compilation}. In
  \bibinfo{booktitle}{\emph{32\textsuperscript{nd} IEEE Computer Security
  Foundations Symposium (CSF~'19)}}.
\newblock
\urldef\tempurl%
\url{https://doi.org/10.1109/CSF.2019.00025}
\showDOI{\tempurl}


\bibitem[Acay et~al\mbox{.}(2021)]%
        {viaduct21}
\bibfield{author}{\bibinfo{person}{Co\c{s}ku Acay}, \bibinfo{person}{Rolph
  Recto}, \bibinfo{person}{Joshua Gancher}, \bibinfo{person}{Andrew~C. Myers},
  {and} \bibinfo{person}{Elaine Shi}.} \bibinfo{year}{2021}\natexlab{}.
\newblock \showarticletitle{Viaduct: An Extensible, Optimizing Compiler for
  Secure Distributed Programs}. In
  \bibinfo{booktitle}{\emph{42\textsuperscript{nd} ACM SIGPLAN Conference on
  Programming Language Design and Implementation (PLDI~'21)}}.
\newblock
\urldef\tempurl%
\url{https://doi.org/10.1145/3453483.3454074}
\showDOI{\tempurl}


\bibitem[Arden et~al\mbox{.}(2012)]%
        {mobileCode12}
\bibfield{author}{\bibinfo{person}{Owen Arden}, \bibinfo{person}{Michael~D.
  George}, \bibinfo{person}{Jed Liu}, \bibinfo{person}{K. Vikram},
  \bibinfo{person}{Aslan Askarov}, {and} \bibinfo{person}{Andrew~C. Myers}.}
  \bibinfo{year}{2012}\natexlab{}.
\newblock \showarticletitle{Sharing Mobile Code Securely with Information Flow
  Control}. In \bibinfo{booktitle}{\emph{33\textsuperscript{rd} IEEE Symposium
  on Security and Privacy (S\&P~'12)}}.
\newblock
\urldef\tempurl%
\url{https://doi.org/10.1109/SP.2012.22}
\showDOI{\tempurl}


\bibitem[Arden et~al\mbox{.}(2015)]%
        {flam15}
\bibfield{author}{\bibinfo{person}{Owen Arden}, \bibinfo{person}{Jed Liu},
  {and} \bibinfo{person}{Andrew~C. Myers}.} \bibinfo{year}{2015}\natexlab{}.
\newblock \showarticletitle{Flow-Limited Authorization}. In
  \bibinfo{booktitle}{\emph{28\textsuperscript{th} IEEE Computer Security
  Foundations Symposium (CSF~'15)}}.
\newblock
\urldef\tempurl%
\url{https://doi.org/10.1109/CSF.2015.42}
\showDOI{\tempurl}


\bibitem[Askarov et~al\mbox{.}(2008)]%
        {AskarovHSS08}
\bibfield{author}{\bibinfo{person}{Aslan Askarov}, \bibinfo{person}{Sebastian
  Hunt}, \bibinfo{person}{Andrei Sabelfeld}, {and} \bibinfo{person}{David
  Sands}.} \bibinfo{year}{2008}\natexlab{}.
\newblock \showarticletitle{Termination-Insensitive Noninterference Leaks More
  Than Just a Bit}. In \bibinfo{booktitle}{\emph{13\textsuperscript{th}
  European Symposium on Research in Computer Security (ESORICS~'08)}}.
\newblock
\urldef\tempurl%
\url{https://doi.org/10.1007/978-3-540-88313-5_22}
\showDOI{\tempurl}


\bibitem[Askarov and Myers(2011)]%
        {AskarovM11}
\bibfield{author}{\bibinfo{person}{Aslan Askarov} {and}
  \bibinfo{person}{Andrew~C. Myers}.} \bibinfo{year}{2011}\natexlab{}.
\newblock \showarticletitle{Attacker Control and Impact for Confidentiality and
  Integrity}.
\newblock \bibinfo{journal}{\emph{Logical Methods in Computer Science (LMCS)}}
  \bibinfo{volume}{7}, \bibinfo{number}{3} (\bibinfo{date}{Sept.}
  \bibinfo{year}{2011}).
\newblock
\urldef\tempurl%
\url{https://doi.org/10.2168/LMCS-7(3:17)2011}
\showDOI{\tempurl}


\bibitem[Askarov and Sabelfeld(2007)]%
        {AskarovS07}
\bibfield{author}{\bibinfo{person}{Aslan Askarov} {and} \bibinfo{person}{Andrei
  Sabelfeld}.} \bibinfo{year}{2007}\natexlab{}.
\newblock \showarticletitle{Gradual Release: Unifying Declassification,
  Encryption and Key Release Policies}. In
  \bibinfo{booktitle}{\emph{28\textsuperscript{th} IEEE Symposium on Security
  and Privacy (S\&P~'07)}}.
\newblock
\urldef\tempurl%
\url{https://doi.org/10.1109/SP.2007.22}
\showDOI{\tempurl}


\bibitem[Bay and Askarov(2020)]%
        {BayA20}
\bibfield{author}{\bibinfo{person}{Johan Bay} {and} \bibinfo{person}{Aslan
  Askarov}.} \bibinfo{year}{2020}\natexlab{}.
\newblock \showarticletitle{Reconciling progress-insensitive noninterference
  and declassification}. In \bibinfo{booktitle}{\emph{33\textsuperscript{rd}
  IEEE Computer Security Foundations Symposium (CSF~'20)}}.
\newblock
\urldef\tempurl%
\url{https://doi.org/10.1109/CSF49147.2020.00015}
\showDOI{\tempurl}


\bibitem[Beutner and Finkbeiner(2022)]%
        {BeutnerF22}
\bibfield{author}{\bibinfo{person}{Raven Beutner} {and} \bibinfo{person}{Bernd
  Finkbeiner}.} \bibinfo{year}{2022}\natexlab{}.
\newblock \showarticletitle{Software Verification of Hyperproperties Beyond
  $k$-Safety}. In \bibinfo{booktitle}{\emph{34\textsuperscript{th}
  International Conference on Computer Aided Verification (CAV~'22)}}.
\newblock
\urldef\tempurl%
\url{https://doi.org/10.1007/978-3-031-13185-1_17}
\showDOI{\tempurl}


\bibitem[Cecchetti([n.\,d.])]%
        {rocq-artifact}
\bibfield{author}{\bibinfo{person}{Ethan Cecchetti}.}
  \bibinfo{year}{[n.\,d.]}\natexlab{}.
\newblock \bibinfo{title}{Nonmalleable Progress Leakage Rocq Proofs}.
\newblock \bibinfo{howpublished}{https://zenodo.org/records/15384760}.
\newblock
\urldef\tempurl%
\url{https://doi.org/10.5281/zenodo.15384760}
\showDOI{\tempurl}


\bibitem[Cecchetti et~al\mbox{.}(2017)]%
        {nmifc17}
\bibfield{author}{\bibinfo{person}{Ethan Cecchetti}, \bibinfo{person}{Andrew~C.
  Myers}, {and} \bibinfo{person}{Owen Arden}.} \bibinfo{year}{2017}\natexlab{}.
\newblock \showarticletitle{Nonmalleable Information Flow Control}. In
  \bibinfo{booktitle}{\emph{24\textsuperscript{th} ACM Conference on Computer
  and Communication Security (CCS~'17)}}.
\newblock
\urldef\tempurl%
\url{https://doi.org/10.1145/3133956.3134054}
\showDOI{\tempurl}


\bibitem[Cecchetti et~al\mbox{.}(2021)]%
        {CecchettiYNM21}
\bibfield{author}{\bibinfo{person}{Ethan Cecchetti}, \bibinfo{person}{Siqiu
  Yao}, \bibinfo{person}{Haobin Ni}, {and} \bibinfo{person}{Andrew~C. Myers}.}
  \bibinfo{year}{2021}\natexlab{}.
\newblock \showarticletitle{Compositional Security for Reentrant Applications}.
  In \bibinfo{booktitle}{\emph{42\textsuperscript{nd} IEEE Symposium on
  Security and Privacy (S\&P~'21)}}.
\newblock
\urldef\tempurl%
\url{https://doi.org/10.1109/SP40001.2021.00084}
\showDOI{\tempurl}


\bibitem[Chong and Myers(2006)]%
        {ChongM06}
\bibfield{author}{\bibinfo{person}{Stephen Chong} {and}
  \bibinfo{person}{Andrew~C. Myers}.} \bibinfo{year}{2006}\natexlab{}.
\newblock \showarticletitle{Decentralized Robustness}. In
  \bibinfo{booktitle}{\emph{19\textsuperscript{th} IEEE Computer Security
  Foundations Workshop (CSFW~'06)}}.
\newblock
\urldef\tempurl%
\url{https://doi.org/10.1109/CSFW.2006.11}
\showDOI{\tempurl}


\bibitem[Chong and Myers(2008)]%
        {ChongM08}
\bibfield{author}{\bibinfo{person}{Stephen Chong} {and}
  \bibinfo{person}{Andrew~C. Myers}.} \bibinfo{year}{2008}\natexlab{}.
\newblock \showarticletitle{End-to-End Enforcement of Erasure and
  Declassification}. In \bibinfo{booktitle}{\emph{21\textsuperscript{st} IEEE
  Computer Security Foundations Symposium (CSF~'08)}}.
\newblock
\urldef\tempurl%
\url{https://doi.org/10.1109/CSF.2008.12}
\showDOI{\tempurl}


\bibitem[Clarkson and Schneider(2010)]%
        {hyperproperties10}
\bibfield{author}{\bibinfo{person}{Michael~R. Clarkson} {and}
  \bibinfo{person}{Fred~B. Schneider}.} \bibinfo{year}{2010}\natexlab{}.
\newblock \showarticletitle{Hyperproperties}.
\newblock \bibinfo{journal}{\emph{Journal of Computer Security ({JCS})}}
  \bibinfo{volume}{18}, \bibinfo{number}{6} (\bibinfo{year}{2010}),
  \bibinfo{pages}{1157--1210}.
\newblock
\urldef\tempurl%
\url{https://doi.org/10.3233/JCS-2009-0393}
\showDOI{\tempurl}


\bibitem[Coenen et~al\mbox{.}(2019)]%
        {CoenenFST19}
\bibfield{author}{\bibinfo{person}{Norine Coenen}, \bibinfo{person}{Bernd
  Finkbeiner}, \bibinfo{person}{C\'{e}sar S\'{a}nchez}, {and}
  \bibinfo{person}{Leander Tentrup}.} \bibinfo{year}{2019}\natexlab{}.
\newblock \showarticletitle{Verifying Hyperliveness}. In
  \bibinfo{booktitle}{\emph{31\textsuperscript{st} International Conference on
  Computer Aided Verification (CAV~'19)}}.
\newblock
\urldef\tempurl%
\url{https://doi.org/10.1007/978-3-030-25540-4_7}
\showDOI{\tempurl}


\bibitem[Farzan and Vandikas(2019)]%
        {FarzanV19}
\bibfield{author}{\bibinfo{person}{Azadeh Farzan} {and}
  \bibinfo{person}{Anthony Vandikas}.} \bibinfo{year}{2019}\natexlab{}.
\newblock \showarticletitle{Automated Hypersafety Verification}. In
  \bibinfo{booktitle}{\emph{31\textsuperscript{st} International Conference on
  Computer Aided Verification (CAV~'19)}}.
\newblock
\urldef\tempurl%
\url{https://doi.org/10.1007/978-3-030-25540-4_11}
\showDOI{\tempurl}


\bibitem[Galatos(2007)]%
        {Galatos07}
\bibfield{author}{\bibinfo{person}{Nikolaos Galatos}.}
  \bibinfo{year}{2007}\natexlab{}.
\newblock \bibinfo{booktitle}{\emph{Residuated Lattices: An Algebraic Glimpse
  at Substructural Logics}}.
\newblock \bibinfo{publisher}{Elsevier Sience}.
\newblock
\showISBNx{0444521410}


\bibitem[Goguen and Meseguer(1982)]%
        {GoguenM82}
\bibfield{author}{\bibinfo{person}{Joseph~A. Goguen} {and}
  \bibinfo{person}{Jos{\'e} Meseguer}.} \bibinfo{year}{1982}\natexlab{}.
\newblock \showarticletitle{Security Policies and Security Models}. In
  \bibinfo{booktitle}{\emph{3\textsuperscript{rd} IEEE Symposium on Security
  and Privacy (S\&P~'82)}}.
\newblock
\urldef\tempurl%
\url{https://doi.org/10.1109/SP.1982.10014}
\showDOI{\tempurl}


\bibitem[Hirsch and Cecchetti(2021)]%
        {HirschC21}
\bibfield{author}{\bibinfo{person}{Andrew~K. Hirsch} {and}
  \bibinfo{person}{Ethan Cecchetti}.} \bibinfo{year}{2021}\natexlab{}.
\newblock \showarticletitle{Giving Semantics to Program-Counter Labels via
  Secure Effects}.
\newblock \bibinfo{journal}{\emph{Proceedings of the ACM on Programming
  Languages}} \bibinfo{volume}{5}, \bibinfo{number}{POPL}, Article
  \bibinfo{articleno}{35} (\bibinfo{date}{Jan.} \bibinfo{year}{2021}),
  \bibinfo{numpages}{29}~pages.
\newblock
\urldef\tempurl%
\url{https://doi.org/10.1145/3434316}
\showDOI{\tempurl}


\bibitem[Lamport and Schneider(2021)]%
        {LamportS21}
\bibfield{author}{\bibinfo{person}{Leslie Lamport} {and}
  \bibinfo{person}{Fred~B. Schneider}.} \bibinfo{year}{2021}\natexlab{}.
\newblock \showarticletitle{Verifying Hyperproperties With {TLA}}. In
  \bibinfo{booktitle}{\emph{34\textsuperscript{th} IEEE Computer Security
  Foundations Symposium (CSF~'21)}}.
\newblock
\urldef\tempurl%
\url{https://doi.org/10.1109/CSF51468.2021.00012}
\showDOI{\tempurl}


\bibitem[Li and Zdancewic(2005)]%
        {LiZ05}
\bibfield{author}{\bibinfo{person}{Peng Li} {and} \bibinfo{person}{Steve
  Zdancewic}.} \bibinfo{year}{2005}\natexlab{}.
\newblock \showarticletitle{Downgrading Policies and Relaxed Noninterference}.
  In \bibinfo{booktitle}{\emph{32\textsuperscript{nd} ACM SIGPLAN Symposium on
  Principles of Programming Languages (POPL~'05)}}.
\newblock
\urldef\tempurl%
\url{https://doi.org/10.1145/1040305.1040319}
\showDOI{\tempurl}


\bibitem[Mantel and Sands(2004)]%
        {MantelS04}
\bibfield{author}{\bibinfo{person}{Heiko Mantel} {and} \bibinfo{person}{David
  Sands}.} \bibinfo{year}{2004}\natexlab{}.
\newblock \showarticletitle{Controlled Declassification Based on Intransitive
  Noninterference}. In \bibinfo{booktitle}{\emph{2\textsuperscript{nd} Asian
  Symposium on Programming Languages and Systems (APLAS~'04)}}.
\newblock
\urldef\tempurl%
\url{https://doi.org/10.1007/978-3-540-30477-7_9}
\showDOI{\tempurl}


\bibitem[McCall et~al\mbox{.}(2023)]%
        {McCallBJ23}
\bibfield{author}{\bibinfo{person}{McKenna McCall}, \bibinfo{person}{Abhishek
  Bichhawat}, {and} \bibinfo{person}{Limin Jia}.}
  \bibinfo{year}{2023}\natexlab{}.
\newblock \showarticletitle{Tainted Secure Multi-Execution to Restrict Attacker
  Influence}. In \bibinfo{booktitle}{\emph{30\textsuperscript{th} ACM
  Conference on Computer and Communication Security (CCS~'23)}}.
\newblock
\urldef\tempurl%
\url{https://doi.org/10.1145/3576915.3623110}
\showDOI{\tempurl}


\bibitem[McLean(1992)]%
        {McLean92}
\bibfield{author}{\bibinfo{person}{John McLean}.}
  \bibinfo{year}{1992}\natexlab{}.
\newblock \showarticletitle{Proving Noninterference and Functional Correctness
  Using Traces}.
\newblock \bibinfo{journal}{\emph{Journal of Computer Security ({JCS})}}
  \bibinfo{volume}{1}, \bibinfo{number}{1} (\bibinfo{date}{Jan.}
  \bibinfo{year}{1992}), \bibinfo{pages}{37--57}.
\newblock
\urldef\tempurl%
\url{https://doi.org/10.3233/JCS-1992-1103}
\showDOI{\tempurl}


\bibitem[Moore et~al\mbox{.}(2012)]%
        {MooreAC12}
\bibfield{author}{\bibinfo{person}{Scott Moore}, \bibinfo{person}{Aslan
  Askarov}, {and} \bibinfo{person}{Stephen Chong}.}
  \bibinfo{year}{2012}\natexlab{}.
\newblock \showarticletitle{Precise Enforcement of Progress-Sensitive
  Security}. In \bibinfo{booktitle}{\emph{19\textsuperscript{th} ACM Conference
  on Computer and Communication Security (CCS~'12)}}.
\newblock
\urldef\tempurl%
\url{https://doi.org/10.1145/2382196.2382289}
\showDOI{\tempurl}


\bibitem[Myers and Liskov(1998)]%
        {MyersL98}
\bibfield{author}{\bibinfo{person}{Andrew~C. Myers} {and}
  \bibinfo{person}{Barbara Liskov}.} \bibinfo{year}{1998}\natexlab{}.
\newblock \showarticletitle{Complete, Safe Information Flow with Decentralized
  Labels}. In \bibinfo{booktitle}{\emph{19\textsuperscript{th} IEEE Symposium
  on Security and Privacy (S\&P~'98)}}.
\newblock
\urldef\tempurl%
\url{https://doi.org/10.1109/SECPRI.1998.674834}
\showDOI{\tempurl}


\bibitem[Myers et~al\mbox{.}(2006)]%
        {MyersSZ06}
\bibfield{author}{\bibinfo{person}{Andrew~C. Myers}, \bibinfo{person}{Andrei
  Sabelfeld}, {and} \bibinfo{person}{Steve Zdancewic}.}
  \bibinfo{year}{2006}\natexlab{}.
\newblock \showarticletitle{Enforcing Robust Declassification and Qualified
  Robustness}.
\newblock \bibinfo{journal}{\emph{Journal of Computer Security ({JCS})}}
  \bibinfo{volume}{14}, \bibinfo{number}{2} (\bibinfo{year}{2006}),
  \bibinfo{pages}{157--196}.
\newblock
\urldef\tempurl%
\url{https://doi.org/10.3233/JCS-2006-14203}
\showDOI{\tempurl}


\bibitem[Nanevski et~al\mbox{.}(2011)]%
        {NanevskiBG11}
\bibfield{author}{\bibinfo{person}{Aleksandar Nanevski},
  \bibinfo{person}{Anindya Banerjee}, {and} \bibinfo{person}{Deepak Garg}.}
  \bibinfo{year}{2011}\natexlab{}.
\newblock \showarticletitle{Verification of Information Flow and Access Control
  Policies with Dependent Types}. In
  \bibinfo{booktitle}{\emph{32\textsuperscript{nd} IEEE Symposium on Security
  and Privacy (S\&P~'11)}}.
\newblock
\urldef\tempurl%
\url{https://doi.org/10.1109/SP.2011.12}
\showDOI{\tempurl}


\bibitem[O'Neill et~al\mbox{.}(2006)]%
        {ONeillCC06}
\bibfield{author}{\bibinfo{person}{Kevin~R. O'Neill},
  \bibinfo{person}{Michael~R. Clarkson}, {and} \bibinfo{person}{Stephen
  Chong}.} \bibinfo{year}{2006}\natexlab{}.
\newblock \showarticletitle{Information-Flow Security for Interactive
  Programs}. In \bibinfo{booktitle}{\emph{19\textsuperscript{th} IEEE Computer
  Security Foundations Workshop (CSFW~'06)}}.
\newblock
\urldef\tempurl%
\url{https://doi.org/10.1109/CSFW.2006.16}
\showDOI{\tempurl}


\bibitem[Pinsky(1995)]%
        {Pinsky95}
\bibfield{author}{\bibinfo{person}{Sylvan Pinsky}.}
  \bibinfo{year}{1995}\natexlab{}.
\newblock \showarticletitle{Absorbing covers and intransitive
  non-interference}. In \bibinfo{booktitle}{\emph{16\textsuperscript{th} IEEE
  Symposium on Security and Privacy (S\&P~'95)}}.
\newblock
\urldef\tempurl%
\url{https://doi.org/10.1109/SECPRI.1995.398926}
\showDOI{\tempurl}


\bibitem[Polikarpova et~al\mbox{.}(2020)]%
        {lifty20}
\bibfield{author}{\bibinfo{person}{Nadia Polikarpova}, \bibinfo{person}{Deian
  Stefan}, \bibinfo{person}{Jean Yang}, \bibinfo{person}{Shachar Itzhaky},
  \bibinfo{person}{Travis Hance}, {and} \bibinfo{person}{Armando
  Solar-Lezama}.} \bibinfo{year}{2020}\natexlab{}.
\newblock \showarticletitle{Liquid Information Flow Control}.
\newblock \bibinfo{journal}{\emph{Proceedings of the ACM on Programming
  Languages}} \bibinfo{volume}{4}, \bibinfo{number}{ICFP}, Article
  \bibinfo{articleno}{105} (\bibinfo{date}{Aug.} \bibinfo{year}{2020}),
  \bibinfo{numpages}{30}~pages.
\newblock
\urldef\tempurl%
\url{https://doi.org/10.1145/3408987}
\showDOI{\tempurl}


\bibitem[{Rocq development team}(2025)]%
        {rocq}
\bibfield{author}{\bibinfo{person}{{Rocq development team}}.}
  \bibinfo{year}{2025}\natexlab{}.
\newblock \bibinfo{booktitle}{\emph{The Rocq Prover}}.
\newblock
\urldef\tempurl%
\url{https://rocq-prover.org/}
\showURL{%
\tempurl}
\newblock
\shownote{Version 8.20.1}.


\bibitem[Roscoe(1995)]%
        {Roscoe95}
\bibfield{author}{\bibinfo{person}{A.W. Roscoe}.}
  \bibinfo{year}{1995}\natexlab{}.
\newblock \showarticletitle{{CSP} and determinism in security modelling}. In
  \bibinfo{booktitle}{\emph{16\textsuperscript{th} IEEE Symposium on Security
  and Privacy (S\&P~'95)}}.
\newblock
\urldef\tempurl%
\url{https://doi.org/10.1109/SECPRI.1995.398927}
\showDOI{\tempurl}


\bibitem[Roscoe and Goldsmith(1999)]%
        {RoscoeG99}
\bibfield{author}{\bibinfo{person}{Andrew~W. Roscoe} {and}
  \bibinfo{person}{Michael~H. Goldsmith}.} \bibinfo{year}{1999}\natexlab{}.
\newblock \showarticletitle{What is Intransitive Noninterference?}. In
  \bibinfo{booktitle}{\emph{12\textsuperscript{th} IEEE Computer Security
  Foundations Workshop (CSFW~'99)}}.
\newblock
\urldef\tempurl%
\url{https://doi.org/10.1109/CSFW.1999.779776}
\showDOI{\tempurl}


\bibitem[Sabelfeld and Myers(2003a)]%
        {SabelfeldM03}
\bibfield{author}{\bibinfo{person}{Andrei Sabelfeld} {and}
  \bibinfo{person}{Andrew~C. Myers}.} \bibinfo{year}{2003}\natexlab{a}.
\newblock \showarticletitle{Language-Based Information-Flow Security}.
\newblock \bibinfo{journal}{\emph{IEEE Journal on Selected Areas in
  Communications}} \bibinfo{volume}{21}, \bibinfo{number}{1}
  (\bibinfo{date}{Jan.} \bibinfo{year}{2003}), \bibinfo{pages}{5--19}.
\newblock
\urldef\tempurl%
\url{https://doi.org/10.1109/JSAC.2002.806121}
\showDOI{\tempurl}


\bibitem[Sabelfeld and Myers(2003b)]%
        {delimitedRelease03}
\bibfield{author}{\bibinfo{person}{Andrei Sabelfeld} {and}
  \bibinfo{person}{Andrew~C. Myers}.} \bibinfo{year}{2003}\natexlab{b}.
\newblock \showarticletitle{A Model for Delimited Information Release}. In
  \bibinfo{booktitle}{\emph{International Symposium on Software Security}}.
\newblock
\urldef\tempurl%
\url{https://doi.org/10.1007/978-3-540-37621-7_9}
\showDOI{\tempurl}


\bibitem[Sabelfeld and Sands(2005)]%
        {SabelfeldS05}
\bibfield{author}{\bibinfo{person}{Andrei Sabelfeld} {and}
  \bibinfo{person}{David Sands}.} \bibinfo{year}{2005}\natexlab{}.
\newblock \showarticletitle{Dimensions and Principles of Declassification}. In
  \bibinfo{booktitle}{\emph{18\textsuperscript{th} IEEE Computer Security
  Foundations Workshop (CSFW~'05)}}.
\newblock
\urldef\tempurl%
\url{https://doi.org/10.1109/CSFW.2005.15}
\showDOI{\tempurl}


\bibitem[Soare(2016)]%
        {Soare16}
\bibfield{author}{\bibinfo{person}{Robert~I. Soare}.}
  \bibinfo{year}{2016}\natexlab{}.
\newblock \bibinfo{booktitle}{\emph{Turing Computability: Theory and
  Applications}}.
\newblock \bibinfo{publisher}{Springer}.
\newblock
\showISBNx{978-3-642-31932-7}
\urldef\tempurl%
\url{https://doi.org/10.1007/978-3-642-31933-4}
\showDOI{\tempurl}


\bibitem[Soloviev et~al\mbox{.}(2024)]%
        {SolovievBG24}
\bibfield{author}{\bibinfo{person}{Matvey Soloviev}, \bibinfo{person}{Musard
  Balliu}, {and} \bibinfo{person}{Roberto Guanciale}.}
  \bibinfo{year}{2024}\natexlab{}.
\newblock \showarticletitle{Security Properties through the Lens of Modal
  Logic}. In \bibinfo{booktitle}{\emph{37\textsuperscript{th} IEEE Computer
  Security Foundations Symposium (CSF~'24)}}.
\newblock
\urldef\tempurl%
\url{https://doi.org/10.1109/CSF61375.2024.00009}
\showDOI{\tempurl}


\bibitem[Sousa and Dillig(2016)]%
        {SousaD16}
\bibfield{author}{\bibinfo{person}{Marcelo Sousa} {and} \bibinfo{person}{Isil
  Dillig}.} \bibinfo{year}{2016}\natexlab{}.
\newblock \showarticletitle{Cartesian Hoare Logic for Verifying $k$-Safety
  Properties}. In \bibinfo{booktitle}{\emph{37\textsuperscript{th} ACM SIGPLAN
  Conference on Programming Language Design and Implementation (PLDI~'16)}}.
\newblock
\urldef\tempurl%
\url{https://doi.org/10.1145/2908080.2908092}
\showDOI{\tempurl}


\bibitem[Stefan et~al\mbox{.}(2011)]%
        {lio11}
\bibfield{author}{\bibinfo{person}{Deian Stefan}, \bibinfo{person}{Alejandro
  Russo}, \bibinfo{person}{John~C. Mitchell}, {and} \bibinfo{person}{David
  Mazi\`{e}res}.} \bibinfo{year}{2011}\natexlab{}.
\newblock \showarticletitle{Flexible Dynamic Information Flow Control in
  {Haskell}}. In \bibinfo{booktitle}{\emph{4\textsuperscript{th} ACM SIGPLAN
  Haskell Symposium (HASKELL~'11)}}.
\newblock
\urldef\tempurl%
\url{https://doi.org/10.1145/2034675.2034688}
\showDOI{\tempurl}


\bibitem[Unno et~al\mbox{.}(2021)]%
        {UnnoTK21}
\bibfield{author}{\bibinfo{person}{Hiroshi Unno}, \bibinfo{person}{Tachio
  Terauchi}, {and} \bibinfo{person}{Eric Koskinen}.}
  \bibinfo{year}{2021}\natexlab{}.
\newblock \showarticletitle{Constraint-Based Relational Verification}. In
  \bibinfo{booktitle}{\emph{33\textsuperscript{rd} International Conference on
  Computer Aided Verification (CAV~'21)}}.
\newblock
\urldef\tempurl%
\url{https://doi.org/10.1007/978-3-030-81685-8_35}
\showDOI{\tempurl}


\bibitem[van~der Meyden(2007)]%
        {vanderMeyden07}
\bibfield{author}{\bibinfo{person}{Ron van~der Meyden}.}
  \bibinfo{year}{2007}\natexlab{}.
\newblock \showarticletitle{What, Indeed, Is Intransitive Noninterference?}. In
  \bibinfo{booktitle}{\emph{12\textsuperscript{th} European Symposium on
  Research in Computer Security (ESORICS~'07)}}.
\newblock
\urldef\tempurl%
\url{https://doi.org/10.1007/978-3-540-74835-9_16}
\showDOI{\tempurl}


\bibitem[Volpano and Smith(1997)]%
        {VolpanoS97}
\bibfield{author}{\bibinfo{person}{Dennis Volpano} {and}
  \bibinfo{person}{Geoffrey Smith}.} \bibinfo{year}{1997}\natexlab{}.
\newblock \showarticletitle{Eliminating Covert Flows with Minimum Typings}. In
  \bibinfo{booktitle}{\emph{10\textsuperscript{th} IEEE Computer Security
  Foundations Workshop (CSFW~'97)}}.
\newblock
\urldef\tempurl%
\url{https://doi.org/10.1109/CSFW.1997.596807}
\showDOI{\tempurl}


\bibitem[Volpano et~al\mbox{.}(1996)]%
        {VolpanoSI96}
\bibfield{author}{\bibinfo{person}{Dennis Volpano}, \bibinfo{person}{Geoffrey
  Smith}, {and} \bibinfo{person}{Cynthia Irvine}.}
  \bibinfo{year}{1996}\natexlab{}.
\newblock \showarticletitle{A Sound Type System for Secure Flow Analysis}.
\newblock \bibinfo{journal}{\emph{Journal of Computer Security ({JCS})}}
  \bibinfo{volume}{4}, \bibinfo{number}{2--3} (\bibinfo{year}{1996}),
  \bibinfo{pages}{167--187}.
\newblock
\urldef\tempurl%
\url{https://doi.org/10.3233/JCS-1996-42-304}
\showDOI{\tempurl}


\bibitem[Waye et~al\mbox{.}(2015)]%
        {WayeBKCR15}
\bibfield{author}{\bibinfo{person}{Lucas Waye}, \bibinfo{person}{Pablo Buiras},
  \bibinfo{person}{Dan King}, \bibinfo{person}{Stephen Chong}, {and}
  \bibinfo{person}{Alejandro Russo}.} \bibinfo{year}{2015}\natexlab{}.
\newblock \showarticletitle{It's My Privilege: Controlling Downgrading in
  {DC}-Labels}. In \bibinfo{booktitle}{\emph{11\textsuperscript{th}
  International Workshop on Security and Trust Management ({STM '15})}}.
\newblock


\bibitem[Zagieboylo et~al\mbox{.}(2019)]%
        {ZagieboyloSM19}
\bibfield{author}{\bibinfo{person}{Drew Zagieboylo}, \bibinfo{person}{G.~Edward
  Suh}, {and} \bibinfo{person}{Andrew~C. Myers}.}
  \bibinfo{year}{2019}\natexlab{}.
\newblock \showarticletitle{Using Information Flow to Design an {ISA} that
  Controls Timing Channels}. In
  \bibinfo{booktitle}{\emph{32\textsuperscript{nd} IEEE Computer Security
  Foundations Symposium (CSF~'19)}}.
\newblock
\urldef\tempurl%
\url{https://doi.org/10.1109/CSF.2019.00026}
\showDOI{\tempurl}


\bibitem[Zdancewic and Myers(2001)]%
        {ZdancewicM01}
\bibfield{author}{\bibinfo{person}{Steve Zdancewic} {and}
  \bibinfo{person}{Andrew~C. Myers}.} \bibinfo{year}{2001}\natexlab{}.
\newblock \showarticletitle{Robust Declassification}. In
  \bibinfo{booktitle}{\emph{14\textsuperscript{th} IEEE Computer Security
  Foundations Workshop (CSFW~'01)}}.
\newblock
\urldef\tempurl%
\url{https://doi.org/10.1109/CSFW.2001.930133}
\showDOI{\tempurl}


\bibitem[Zdancewic and Myers(2003)]%
        {ZdancewicM03}
\bibfield{author}{\bibinfo{person}{Steve Zdancewic} {and}
  \bibinfo{person}{Andrew~C. Myers}.} \bibinfo{year}{2003}\natexlab{}.
\newblock \showarticletitle{Observational Determinism for Concurrent Program
  Security}.
\newblock
\urldef\tempurl%
\url{https://doi.org/10.1109/CSFW.2003.1212703}
\showDOI{\tempurl}


\bibitem[Zeldovich et~al\mbox{.}(2011)]%
        {histar11}
\bibfield{author}{\bibinfo{person}{Nickolai Zeldovich}, \bibinfo{person}{Silas
  Boyd-Wickizer}, \bibinfo{person}{Eddie Kohler}, {and} \bibinfo{person}{David
  Mazi\`{e}res}.} \bibinfo{year}{2011}\natexlab{}.
\newblock \showarticletitle{Making Information Flow Explicit in {HiStar}}.
\newblock \bibinfo{journal}{\emph{Commun. ACM}} \bibinfo{volume}{54},
  \bibinfo{number}{11} (\bibinfo{date}{Nov.} \bibinfo{year}{2011}),
  \bibinfo{pages}{93--101}.
\newblock
\urldef\tempurl%
\url{https://doi.org/10.1145/2018396.2018419}
\showDOI{\tempurl}


\end{thebibliography}
